\documentclass[10pt,onecolumn]{revtex4-2} 
\usepackage[dvips]{color}
\usepackage{epsfig}
\usepackage{amssymb}
\usepackage{latexsym}
\usepackage{graphicx}
\begin{document}

\title{Renormalization of $(2+1)D$ scalar Weyl spinors interactions on lattices\\ using the Clifford groups}
\author{Sadataka Furui        }
\affiliation{(Formerly) Graduate School of Science and Engineering, Teikyo University, Utsunomiya, 320 Japan }
 \email{furui@umb.teikyo-u.ac.jp      }
\date{\today }

\begin{abstract}
We consider symplectic quaternions instead of unitary spinors sitting on a lattice, and calculate the fixed point Wilson action
on a finite $2D$ plane expanded by $u_1 a e_1+ u_2 a e_2$ and on two $2D$ planes separated by $a e_1\wedge e_2$. Only the nearlest neighbor interactions are considered. Following Migdal and Kadanoff, we perform
the renormalization of the Wilson action by making the lattice spacing $(\frac{1}{2})^{h}a$ $(h=0,\cdots,11  )$,
in order to simulate bosonic and solitonic phonon propagation in materials. 
Renormalization group method of Benfatto and Gallavotti for $(2+1)D$ scalar $\phi^4$ system for sound propagation in Fermi sea is applied and feasibility of numerical simulation is discussed.

\end{abstract}
\maketitle
\section{ Introduction}
In \cite{SF21}, an outline of performing a simulation of phonetic solitons propagating on $(2+1)D$ plane was proposed.
We start from $4\times 4$ lattices surrounded by Clifford pairs, and by adopting the renormalization group method, calculate the correlations of ultra-sonic phonetic solitons.

A main difference from standard methods is on each lattice site quatrernions following symplectic groups are sitting., not spin following unitary groups.

We considered the fixed point Wilson actions in one loop adopted by deGrand et al.\cite{DGHHN95}. They considered in the $(3+1)D$ lattice, 28 Fixed point (FP) actions of length less than or equal to 8 lattice lengths. We classify the FP actions to $Loop c$ which consist of loops on one $2D$  plane expanded by $e_1$ and $e_2$, and $Loop d$ which consist of loops on two parallel $2D$ planes connected by two links in the direction $e_1\wedge e_2$ and in the direction $e_2\wedge e_1$.
The $Loop 1,2,5,6,11,12,18$ and $28$ belong to the $Loop C$ and the $Loop 3,4,7,8,9,10,13,14,15,16,17$ and $26,27$ belong to the $Loop D$. $Loop 19,\cdots,25$ are irrelevant in $(2+1)D$.

In abstract graph theory of Luescher\cite{Luescher86} a loop in a graph is a non-empty subset of lines with the property that there exists a sequence $v_1,\cdots,v_N$ of pairwise different vertices and a labelling $\ell_1,\cdots,\ell_N$ of the lines in a loop, such that $v_k,v_{k+1}$ are the end points of $\ell_k$ $(k=1,\cdots, N-1)$ and $v_N,v_1$ are end points of $\ell_N$. On every loop there are two orientations, and a crossing of two loops produces a new vertex. 

\begin{figure}[h]
\begin{minipage}[b]{0.47\linewidth}
\begin{center}
\includegraphics[width=3.5cm,angle=0,clip]{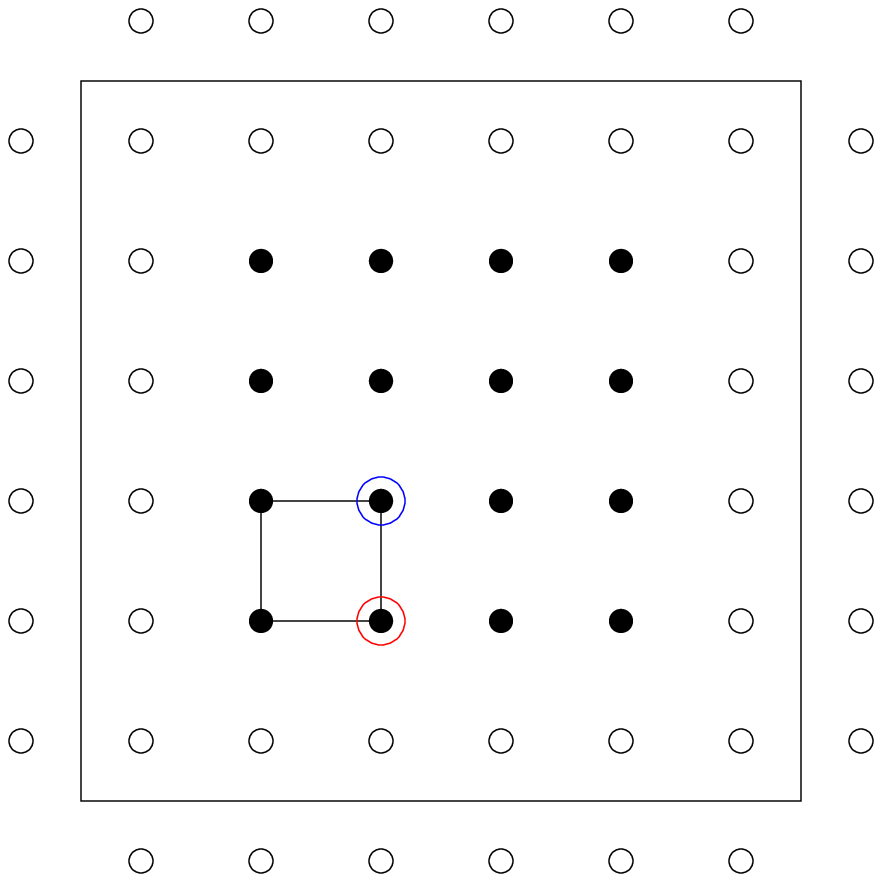} 
\end{center}
\end{minipage}
\hfill
\begin{minipage}[b]{0.47\linewidth}
\begin{center}
\includegraphics[width=3.5cm,angle=0,clip]{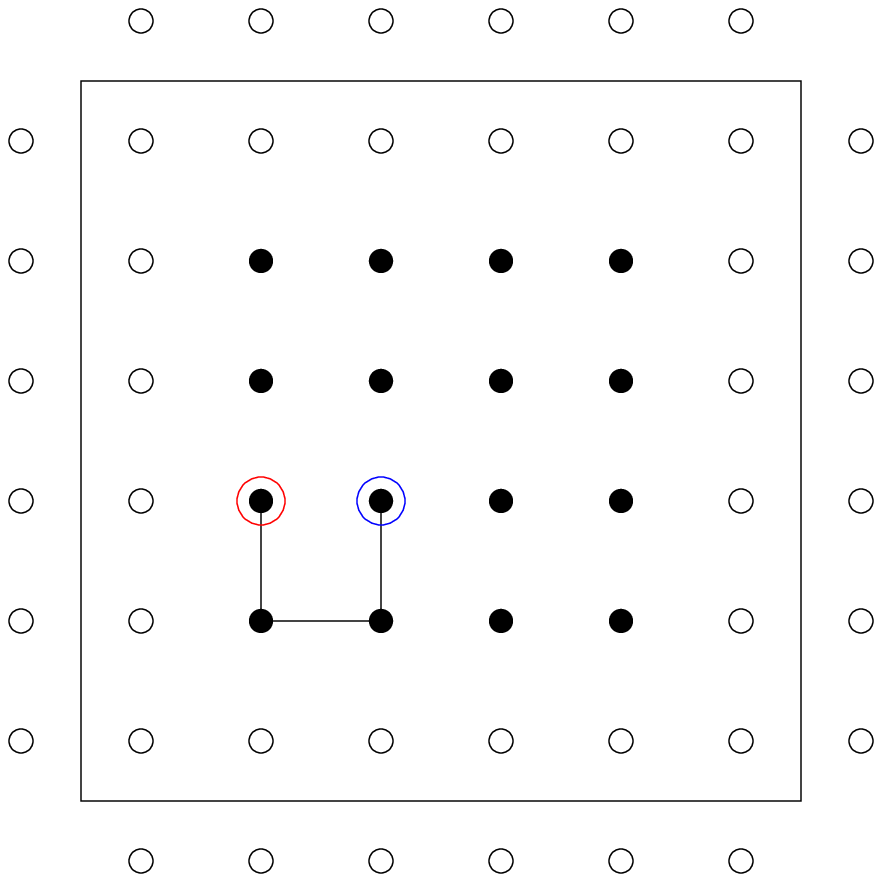} 
\end{center}
\end{minipage}
\caption{ Loop26 (left) and Loop27 (right). }
\label{L2627}
\end{figure}

In this sence, the $Loop28$ is not a proper loop. In \cite{DGHHN95}, the loop did not play important roles, and we also observed that the eigenvalues of the action are large. We did not consider the $Loop 26$ and $27$ which are shown in Fig.\ref{L2627} since eigenvalues were large, but we consider these loops in $(2+1)D$ space, since they are proper. 

Depending on the direction of the link between two planes, we consider paths  

$Loop 26 \alpha$: ${\bf u}\to{\bf u}+a e_1\to$ ${\bf u}+a e_1+a e_2\to$ ${\bf u}+a e_1+a e_2+ a e_1\wedge e_2\to$ ${\bf u}+a e_1+a e_1\wedge e_2\to$  ${\bf u}+a e_1\to$ ${\bf u}+a e_1+a e_2\to$ ${\bf u}+a e_2\to$ ${\bf u}$, 
and

$Loop 26\beta$: ${\bf u}\to{\bf u}+a e_1\to$ ${\bf u}+a e_1+a e_2\to$ ${\bf u}+a e_1+a e_2+ a e_2\wedge e_1\to$ ${\bf u}+a e_1+a e_2\wedge e_1\to$  ${\bf u}+a e_1\to$ ${\bf u}+a e_1+a e_2\to$ ${\bf u}+a e_2\to$ ${\bf u}$.


 Similarly the paths
 
  $Loop27 \alpha$: ${\bf u}\to {\bf u}+a e_1\to$ ${\bf u}+a e_1+a e_2\to$ ${\bf u}+a e_1+a e_2+a e_1\wedge e_2\to$ ${\bf u}+a e_1+a e_1\wedge e_2\to$  ${\bf u}+a e_1\wedge e_2\to$ ${\bf u}+a e_2+a e_1\wedge e_2\to$ ${\bf u}+a e_2\to$ ${\bf u}$,
and
   
$Loop27\beta$: ${\bf u}\to {\bf u}+a e_1\to$ ${\bf u}+a e_1+a e_2\to$ ${\bf u}+a e_1+a e_2+a e_2\wedge e_1\to$ ${\bf u}+a e_1+a e_2\wedge e_1\to$  ${\bf u}+a e_2\wedge e_1\to$ ${\bf u}+a e_2+a e_2\wedge e_1\to$ ${\bf u}+a e_2\to$ ${\bf u}$. 

 In the two loops $\alpha$ and $\beta$, the blue circle and the red circle are to be interchanged. 
 
  The eigenvalues of loops have dependence on the direction of paths. Difference of eigenvalues between $Loop26\alpha$ and $Loop26\beta$ is larger than that of $Loop27\alpha$ and $Loop27\beta$. Their eigenvalues are about the same as those of $Loop28$. 
  
 The dependence of eigenvalues on the direction of $e_1\wedge e_2$ shows a presence of time reversal symmetric but rotational symmetry breaking phase. 
 
We evaluate Wilson's optimum plaquet actions by making a linear combination of eigenvalues of selected FP actions.
 
 The structure of this presentation is as follows.
 In Sec. II, we compare eigenvalues $\chi(L^{(h)})$ of $Loop C$ and $Loop D$ for a lattice spacing $a$ used in \cite{SF20} and $a/2$.
 The similar analysis for the trace of link variables are given in Sec. III.
 In Sec. IV, a perspective of the renormalization group analysis using supercomputer  are given.
 
 \section{Lattice spacing dependence of eigenvalues of plaquettes}
 
 We consider the case in which the spacing between the lattice $\Delta x=\frac{e_i}{4}$ which is called $Loop C$ and they are $\Delta x=\frac{e_i}{8}$ which are called $Loop D$.

\subsection{Paths on one 2$D$ plane expanded by $e_1$ and $e_2$}
The $Loop 1$ consists of 4 sides of a square, whose eigenvalue is the smallest among the FP actions. We characterized the path by ${\bf u}\to$ ${\bf u}+\frac{1}{4}e_1 \to$ ${\bf u}+\frac{e_1}{4}+\frac{e_2}{4}\to$ ${\bf u}+\frac{e_2}{4}\to$ $\bf u$ where ${\bf u}$ are mesh points $(u_1 e_1+u_2 e_2)$, $0\leq u_1,u_2\leq 3$ ($u_i\in {\bf Z}$). We call the path $Loop1c$.

We compare eigenvalues of the path of  ${\bf u}\to$ ${\bf u}+\frac{1}{8}e_1 \to$ ${\bf u}+\frac{e_1}{8}+\frac{e_2}{8}\to$ ${\bf u}+\frac{e_2}{8}\to$ ${\bf u}-\frac{1}{8}e_2\to$ $\bf u$, which we call $Loop1d$.

\begin{figure*}[htb]
\begin{minipage}[b]{0.47\linewidth}
\begin{center}
\includegraphics[width=6cm,angle=0,clip]{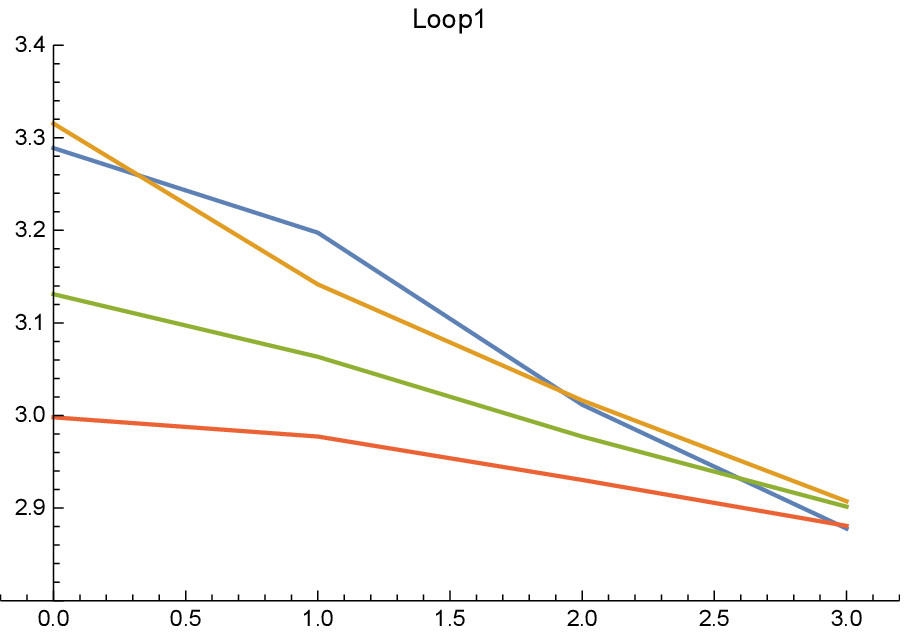} 
\end{center}
\end{minipage}
\hfill
\begin{minipage}[b]{0.47\linewidth}
\begin{center}
\includegraphics[width=6cm,angle=0,clip]{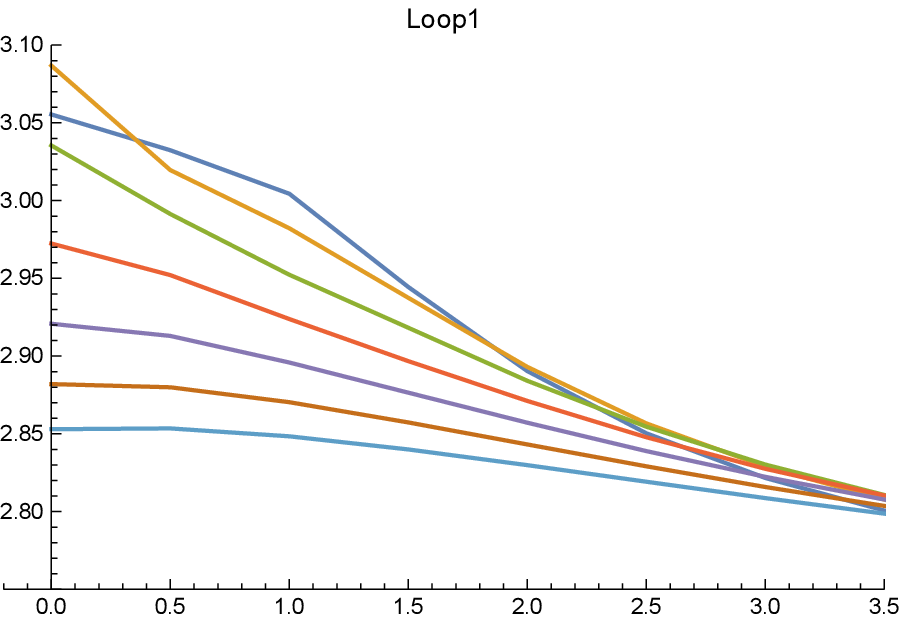} 
\end{center}
\end{minipage}
\caption{ Absolute values of eigenvalues for fixed $u_1$ in Loop1c(left) and in Loop1d (right). }
\label{L1}
\end{figure*}

In the left figure of Fig.\ref{L1}, eigenvalues of $u_1=0 (blue), 1(orange), 2(green), 3(red)$ as a function of $u_2$ are plotted, and in the right figure of Fig.\ref{L1}, eigenvalues of $u_1=0, \frac{1}{2}, 1, \frac{3}{2}, 2, \frac{5}{2}, 3$ as a function of $u_2$ are plotted. When the lattice spacing is smaller, eigenvalues are smaller.

\begin{figure*}[htb]
\begin{minipage}[b]{0.47\linewidth}
\begin{center}
\includegraphics[width=6cm,angle=0,clip]{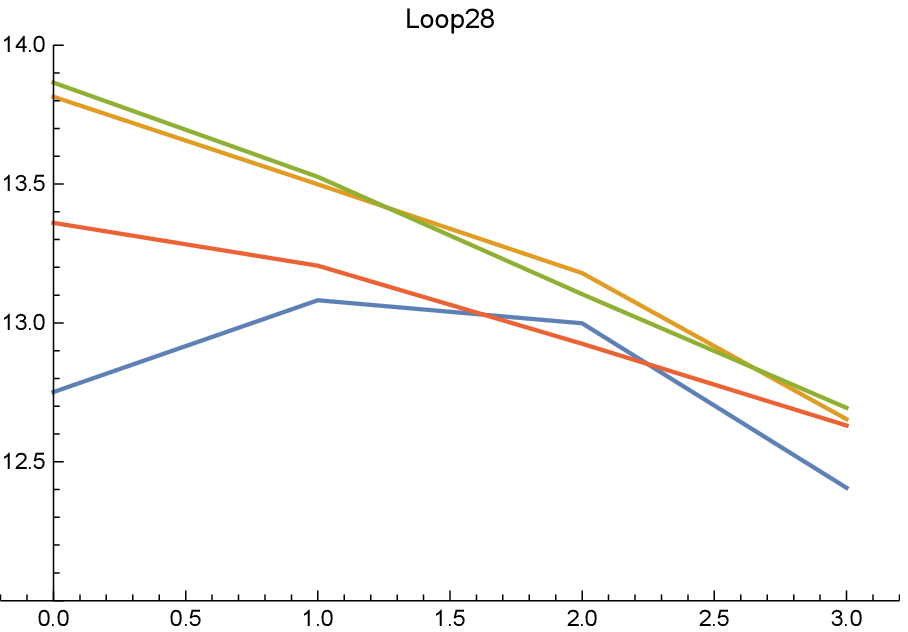} 
\end{center}
\end{minipage}
\hfill
\begin{minipage}[b]{0.47\linewidth}
\begin{center}
\includegraphics[width=6cm,angle=0,clip]{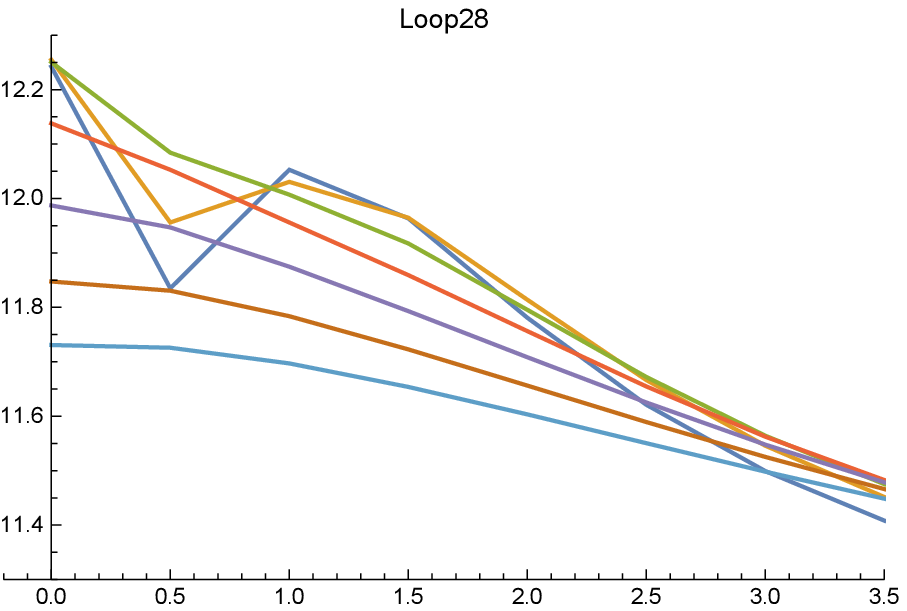} 
\end{center}
\end{minipage}
\caption{ Absolute values of eigenvalues for fixed $u_1$ in Loop28c(left) and in Loop28d(right). }
\label{L28}
\end{figure*}

\begin{figure*}[htb]
\begin{minipage}[b]{0.47\linewidth}
\begin{center}
\includegraphics[width=6cm,angle=0,clip]{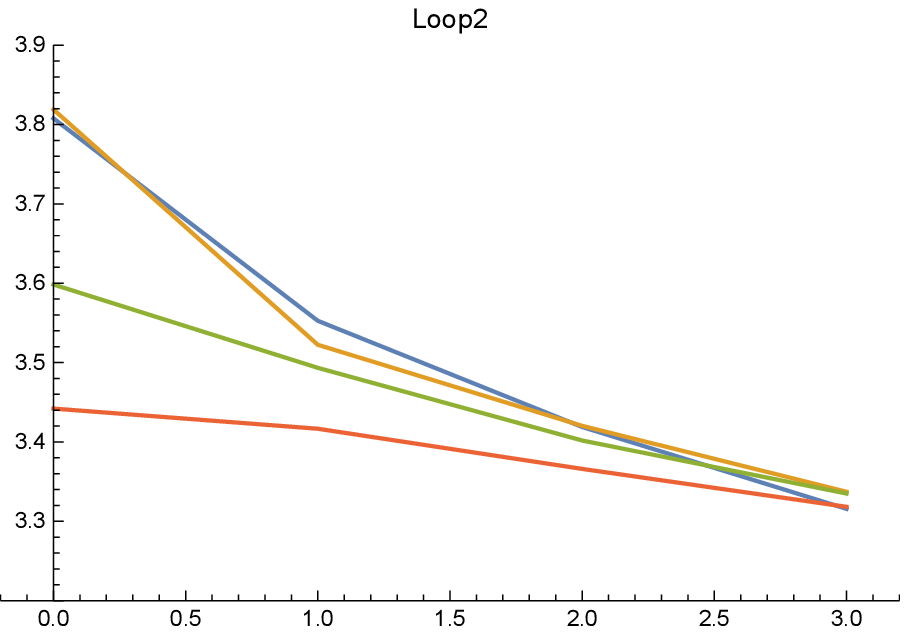} 
\end{center}
\end{minipage}
\hfill
\begin{minipage}[b]{0.47\linewidth}
\begin{center}
\includegraphics[width=6cm,angle=0,clip]{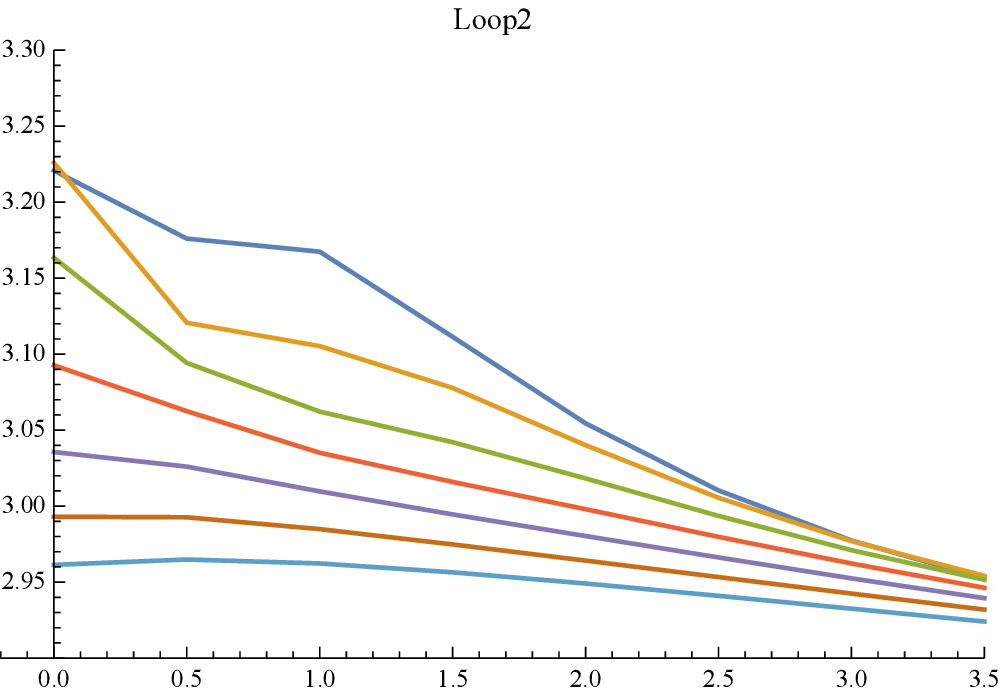} 
\end{center}
\end{minipage}
\caption{ Absolute values of eigenvalues for a fixed $u_1$ in Loop2c(left) and in Loop2d(right) }
\label{L2}
\end{figure*}

The eigenvalues of $Loop28$ are shown in Fig.\ref{L28}.
The loop is not proper in the sence of Luescher\cite{Luescher86}. Fluctuations are large for small ${\bf u}$ or in infrared regions.
\newpage
The eigenvalues of $Loop2$ are shown in Fig.\ref{L2}.
Eigenvalues of the smaller lattice spacing $Loop2d$ is smaller than those of $Loops2c$.

\begin{figure*}[htb]
\begin{minipage}[b]{0.47\linewidth}
\begin{center}
\includegraphics[width=6cm,angle=0,clip]{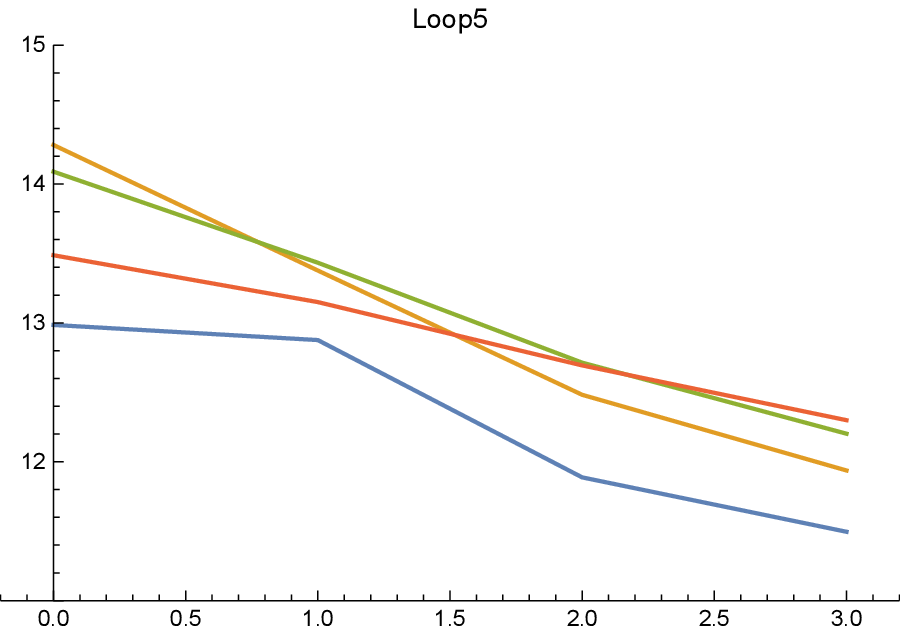} 
\end{center}
\end{minipage}
\hfill
\begin{minipage}[b]{0.47\linewidth}
\begin{center}
\includegraphics[width=6cm,angle=0,clip]{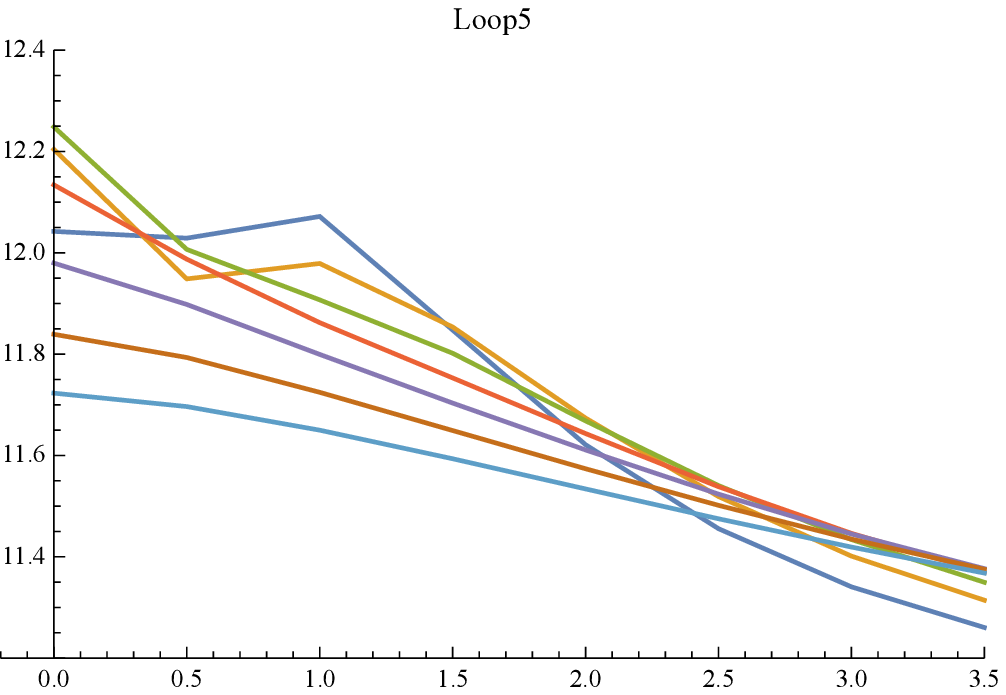} 
\end{center}
\end{minipage}
\caption{ Absolute values of eigenvalues for a fixed $u_1$ in Loop5c(left) and in Loop5d(right) .}
\label{L5}
\end{figure*}

The eigenvalues of $Loop5$ are shown in Fig.\ref{L5}. 
The $Loop5$ has a bending point of the path at the center of the loop. It produces a large eigenvalues for small $u_1$.

\begin{figure*}[htb]
\begin{minipage}[b]{0.47\linewidth}
\begin{center}
\includegraphics[width=6cm,angle=0,clip]{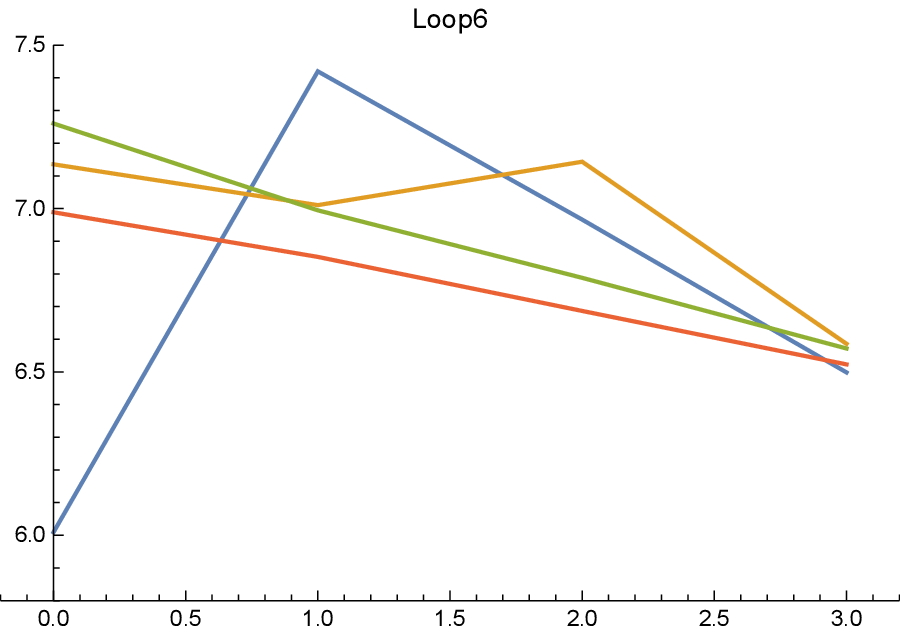} 
\end{center}
\end{minipage}
\hfill
\begin{minipage}[b]{0.47\linewidth}
\begin{center}
\includegraphics[width=6cm,angle=0,clip]{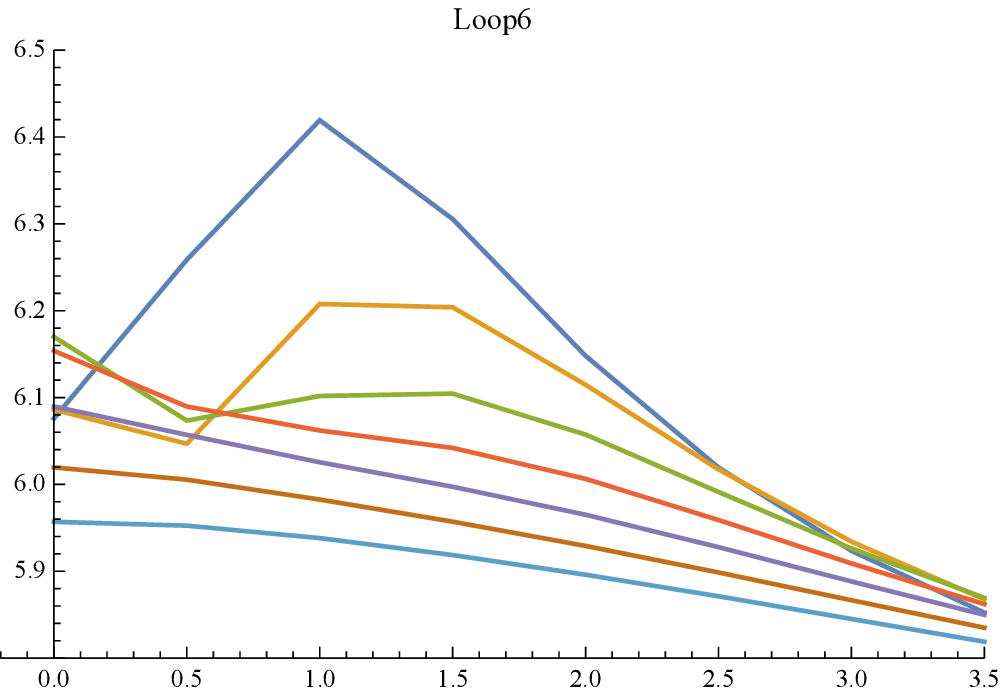} 
\end{center}
\end{minipage}
\caption{ Absolute values of eigenvalues for a fixed $u_1$ in Loop6c (left) and in Loop6d (right).}
\label{L6}
\end{figure*}

\newpage
The eigenvalues of $Loop6$ are shown in Fig.\ref{L6}.
 They have ${\bf u}$ dependence in the infrared region.

\begin{figure*}[htb]
\begin{minipage}[b]{0.47\linewidth}
\begin{center}
\includegraphics[width=6cm,angle=0,clip]{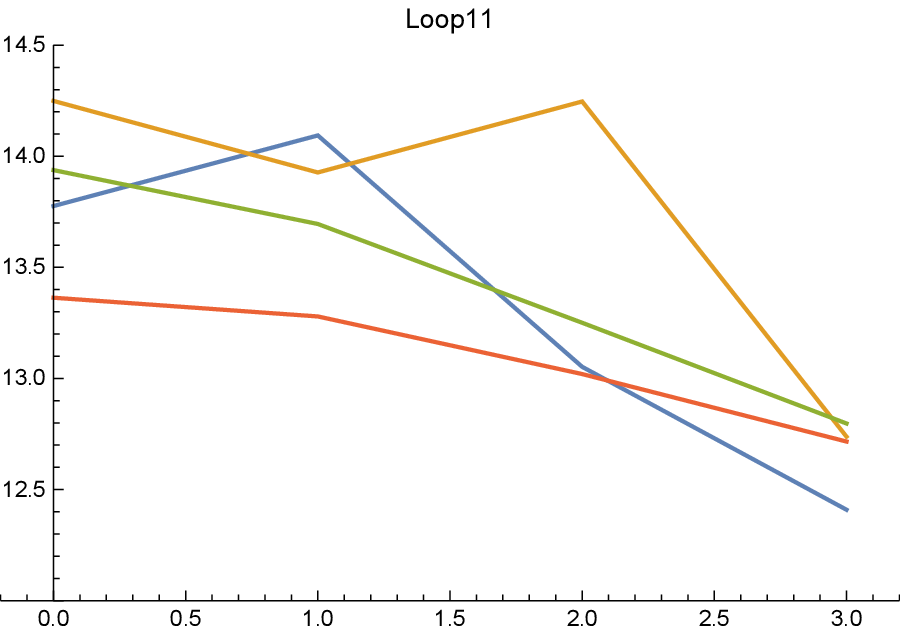} 
\end{center}
\end{minipage}
\hfill
\begin{minipage}[b]{0.47\linewidth}
\begin{center}
\includegraphics[width=6cm,angle=0,clip]{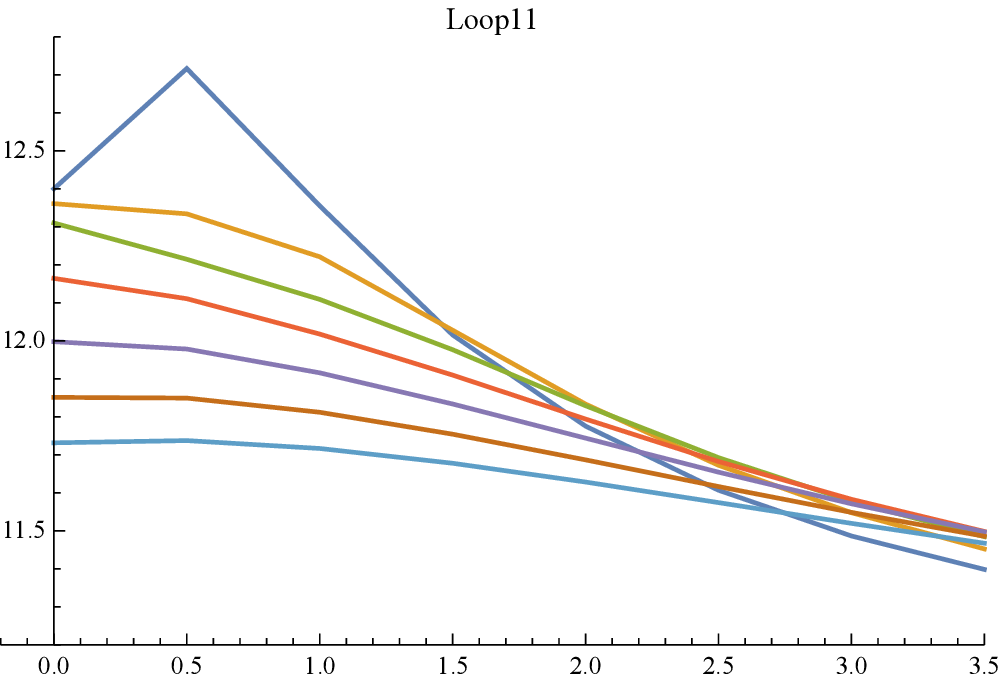} 
\end{center}
\end{minipage}
\caption{ Absolute values of eigenvalues for a fixed $u_1$in Loop11c (left) and in Loop11d (right).}
\label{L11}
\end{figure*}

The eigenvalues of $Loop11$ are shown in Fig.\ref{L11}.
The $Loop11$ has a bending due to presence of overlapping links in the center.

\begin{figure*}[htb]
\begin{minipage}[b]{0.47\linewidth}
\begin{center}
\includegraphics[width=6cm,angle=0,clip]{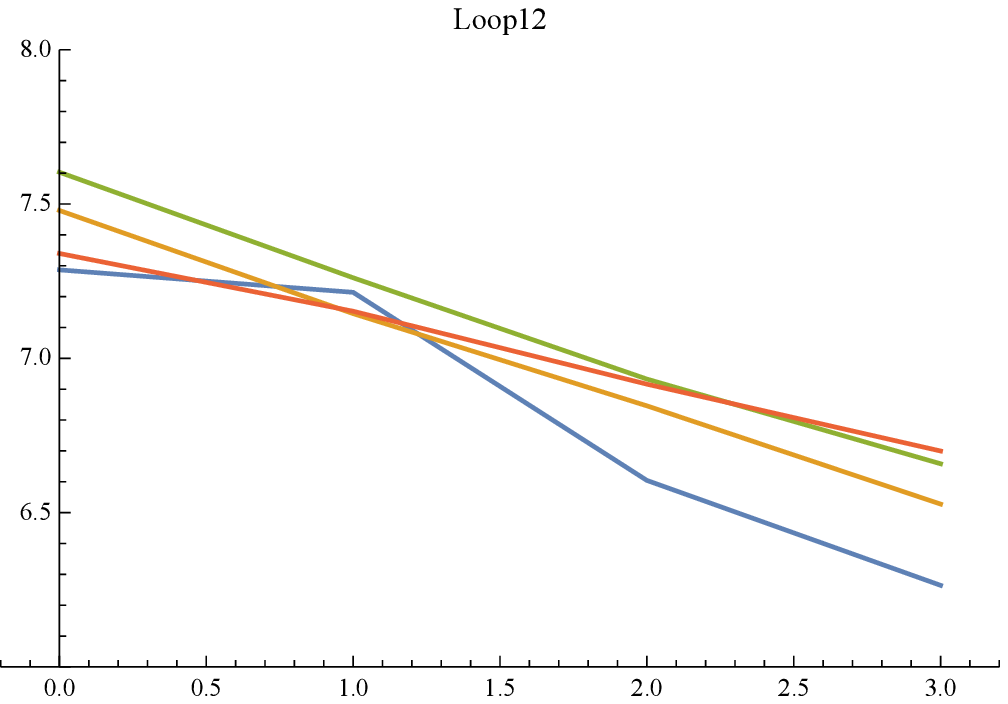} 
\end{center}
\end{minipage}
\hfill
\begin{minipage}[b]{0.47\linewidth}
\begin{center}
\includegraphics[width=6cm,angle=0,clip]{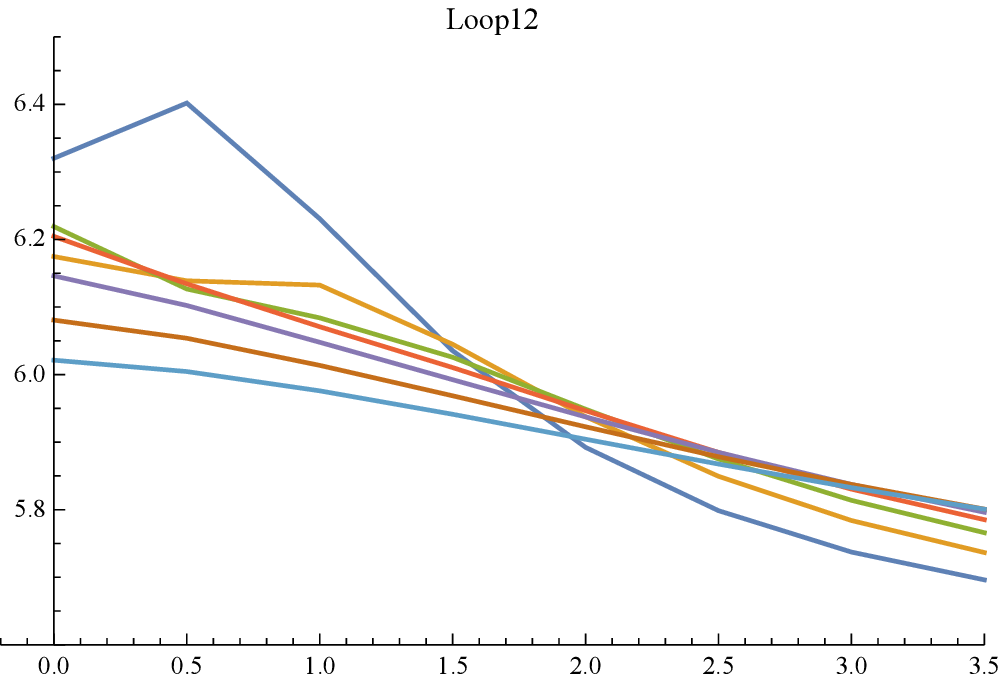} 
\end{center}
\end{minipage}
\caption{ Absolute values of eigenvalues for a fixed $u_1$ in Loop12c (left) and in Loop12d(right). .}
\label{L12}
\end{figure*}

The eigenvalues of $Loop12$ are shown in Fig.\ref{L12}, 
The $Loop12$ is similar to the $Loop5$, but there is a self-crossing of two long links at the center. The  $\bf u$ dependence of eigenvalues of $Loop5$ and $Loop12$ are similar, but absolute values of eigenvalues of $Loop12$ are smaller, due to presence of long links. In the $(3+1)D$ lattice simulation of \cite{DGHHN95},
the $Loop5$ is more effective than $Loop12$.

\begin{figure*}[htb]
\begin{minipage}[b]{0.47\linewidth}
\begin{center}
\includegraphics[width=6cm,angle=0,clip]{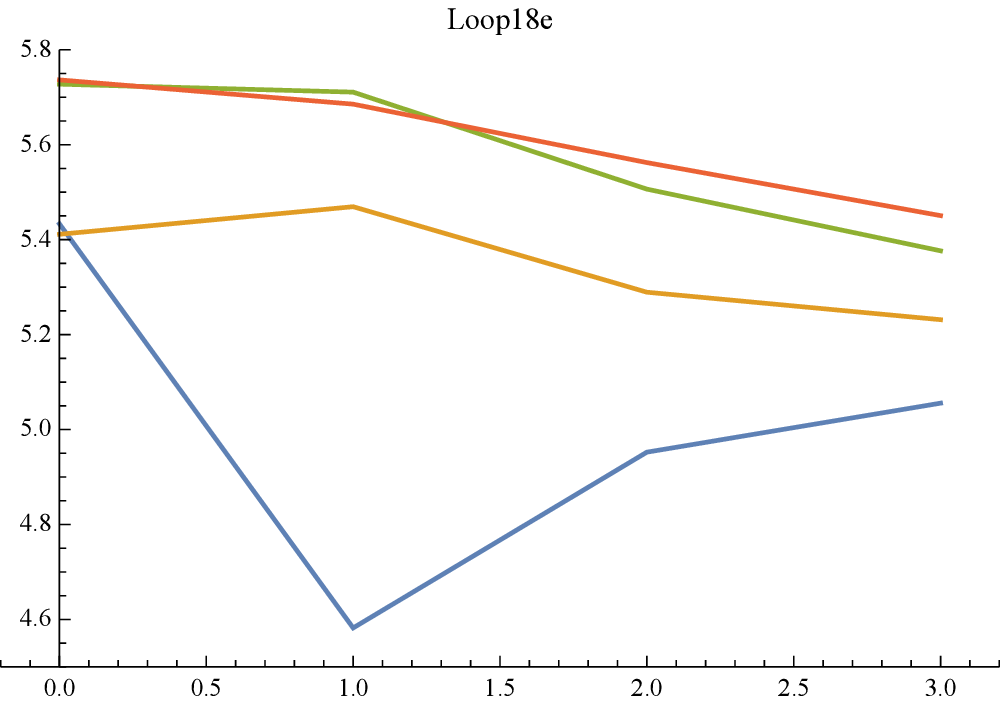}%
\end{center}
\end{minipage}
\hfill
\begin{minipage}[b]{0.47\linewidth}
\begin{center}
\includegraphics[width=6cm,angle=0,clip]{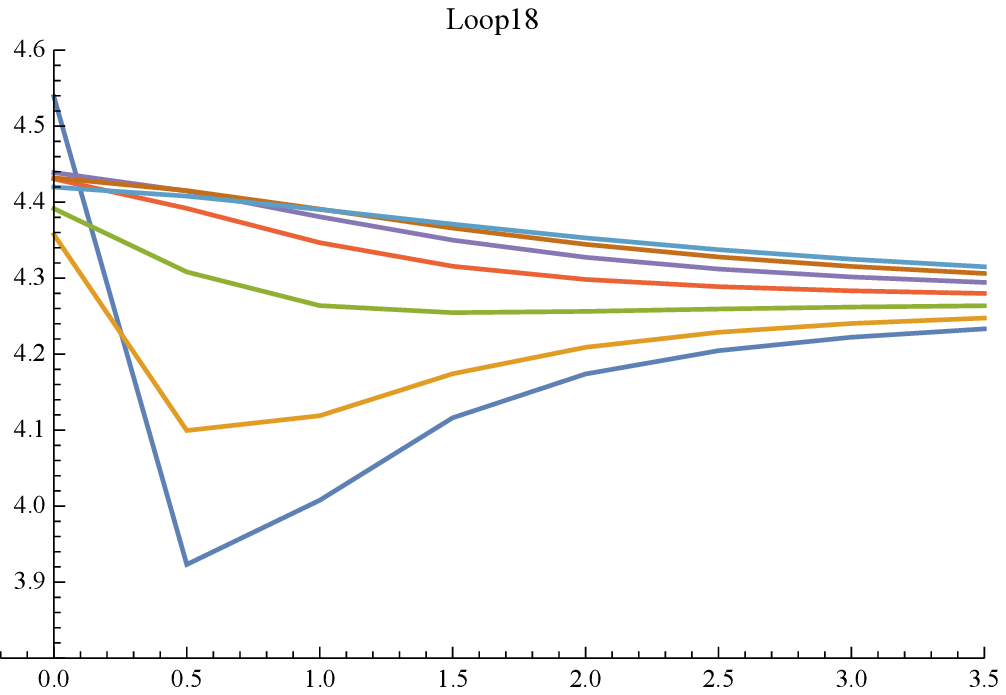}
\end{center}
\end{minipage}
\caption{ Absolute values of eigenvalues for a fixed $u_1$ in Loop18c (left) and in Loop18d (right). }
\label{L18}
\end{figure*}

In \cite{SF21}, the starting point of the$Loop18$ was chosen to be same as the $Loop1$, and only the scale was changed. In this paper we replace the verices as those of the paper \cite{DGHHN95},
\[
L18[u_1,u_2]=t1[-\frac{1}{4},u_1+\frac{1}{4},u_2]\times t2[-\frac{1}{2},u_1+\frac{1}{4},u_2+\frac{1}{2}]\times t1[-\frac{1}{2},u_1+\frac{1}{4},u_2+\frac{1}{2}]\times
t2[\frac{1}{2},u_1+\frac{1}{4},u_2]\times t1[\frac{1}{4},u_1,u_2].
\]
The eigenvalues of the new $Loop18$ and its lattice scalings halved are shown in Fig.\ref{L18}.
The $Loop18$ contains long straight links parallel to $e_2$.  

\subsection{Paths on two planes connected by $e_1\wedge e_2$}
When there are links between two $2D$ planes, a proplem of choice of scale of length between the two 2D planes appears. We leave it as a future study, and calculate eigenvalues using the same lattice spacing between the 2D planes and on the 2D plane.

The paths on two planes we considered in \cite{SF20} were restricted to $\alpha$ type. Since eigenvalues in the large $\bf u$ region are almost independent of $\alpha$ and $\beta$, we restrict loops to be $\alpha$ type and consider $\beta$ type near the final stage of MonteCarlo simulation.

The path of $Loop 3c$ consists of ${\bf u}\to {\bf u}+\frac{1}{4}e_1\to$ ${\bf u}+\frac{1}{4}e_1+\frac{1}{4}e_2\to$ ${\bf u}+\frac{1}{4}e_1+\frac{1}{4}e_2+\frac{1}{4}e_1\wedge e_2\to$ ${\bf u}+\frac{1}{4}e_1+\frac{1}{4}e_1\wedge e_2\to$ ${\bf u}+\frac{1}{4}e_1\wedge e_2\to$ ${\bf u}$.

The path of $Loop 3d$ consists of ${\bf u}\to {\bf u}+\frac{1}{8}e_1\to$ ${\bf u}+\frac{1}{8}e_1 +\frac{1}{8}e_2\to$  ${\bf u}+\frac{1}{8}e_1+\frac{1}{8}e_2+\frac{1}{8}e_1\wedge e_2\to$ ${\bf u}+\frac{1}{8}e_1+\frac{1}{8}e_1\wedge e_2\to$ ${\bf u}+\frac{1}{8}e_1\wedge e_2\to$ ${\bf u}$.

\begin{figure*}[htb]
\begin{minipage}[b]{0.47\linewidth}
\begin{center}
\includegraphics[width=6cm,angle=0,clip]{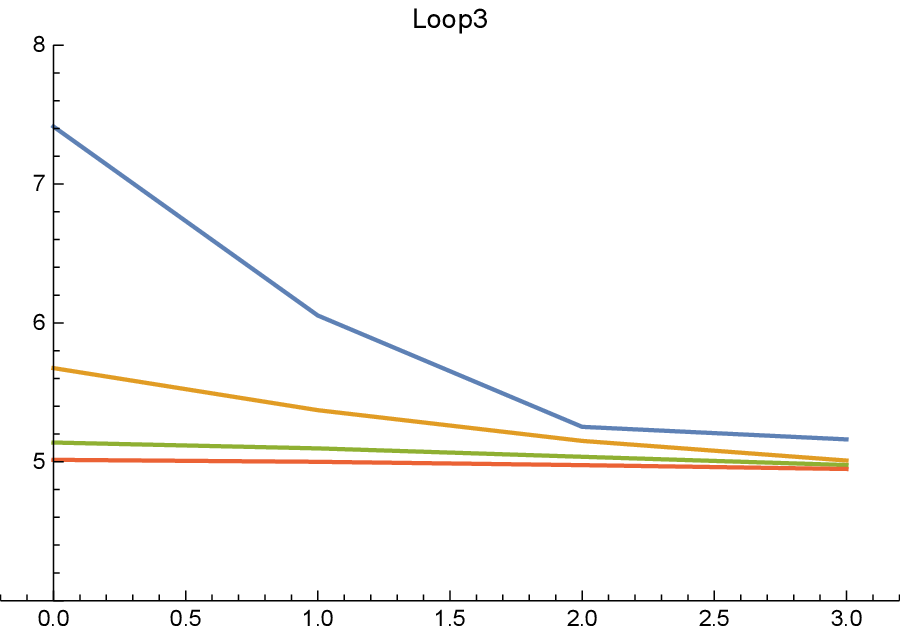} 
\end{center}
\end{minipage}
\begin{minipage}[b]{0.47\linewidth}
\begin{center}
\includegraphics[width=6cm,angle=0,clip]{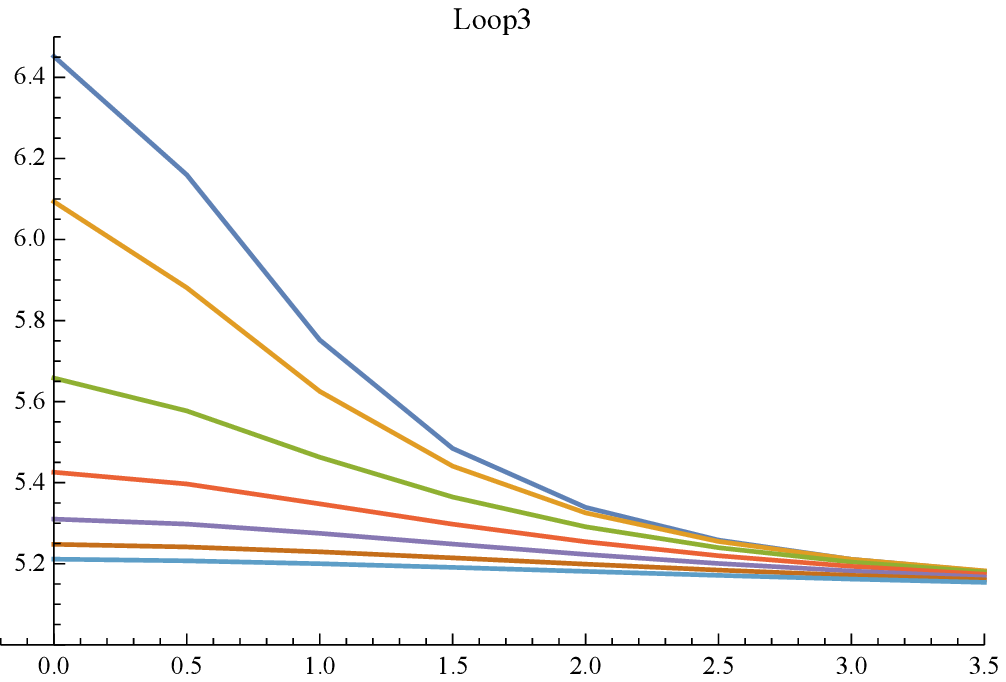} 
\end{center}
\end{minipage}
\caption{ Absolute values of eigenvalues for a fixed $u_1$ in Loop3c (left) and in Loop3d(right). }
\label{L3}
\end{figure*}

The eigenvalues of $Loop 3$ are shown in Fig.\ref{L3} .
We observe the eigenvalues of action of $Loop 3$ is roughly about twice of those of $Loop1$.

\begin{figure*}[htb]
\begin{minipage}[b]{0.47\linewidth}
\begin{center}
\includegraphics[width=6cm,angle=0,clip]{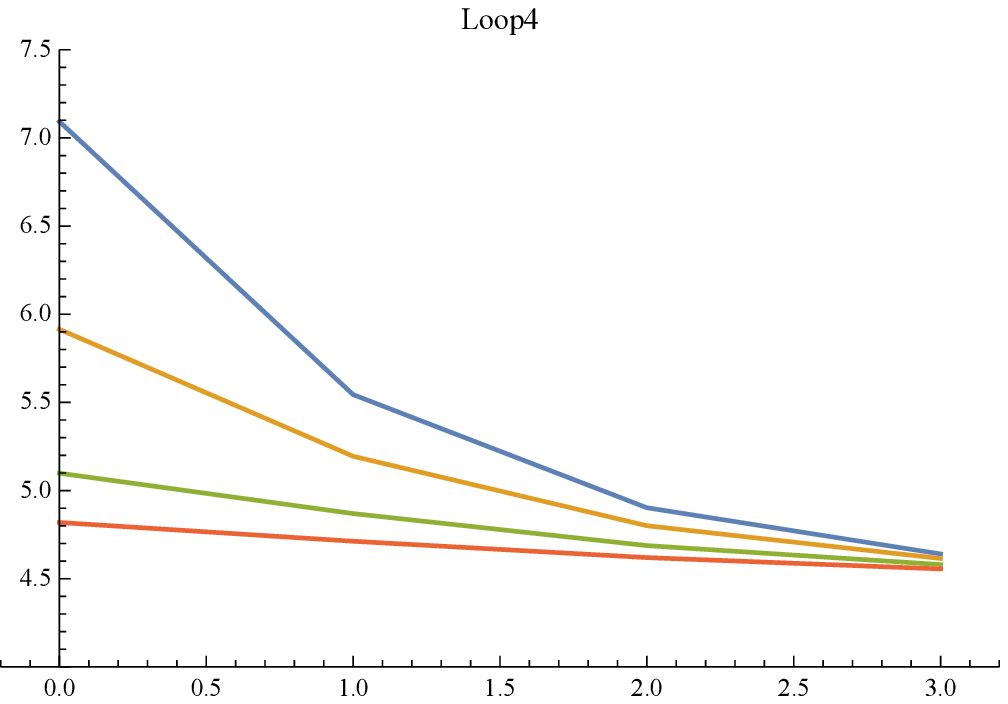} 
\end{center}
\end{minipage}
\begin{minipage}[b]{0.47\linewidth}
\begin{center}
\includegraphics[width=6cm,angle=0,clip]{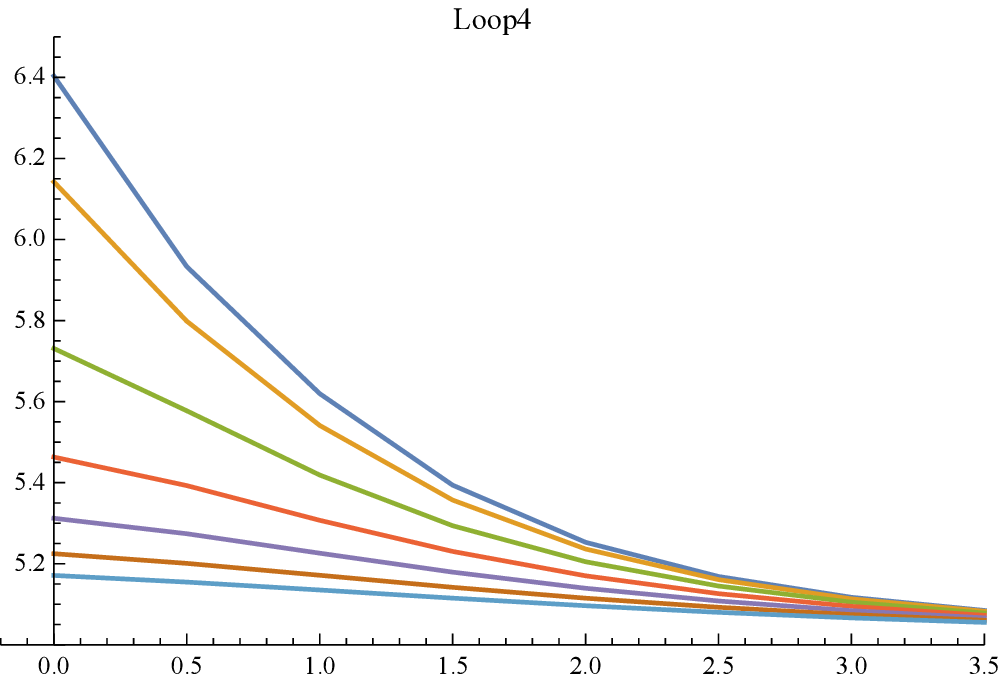} 
\end{center}
\end{minipage}
\caption{ Absolute values of eigenvalues for a fixed $u_1$ in Loop4c (left) and in Loop4d(right).  }
\label{L4}
\end{figure*}

The eigenvalues of $Loop 4$ are shown in Fig.\ref{L4}.

\begin{figure*}[htb]
\begin{minipage}[b]{0.47\linewidth}
\begin{center}
\includegraphics[width=6cm,angle=0,clip]{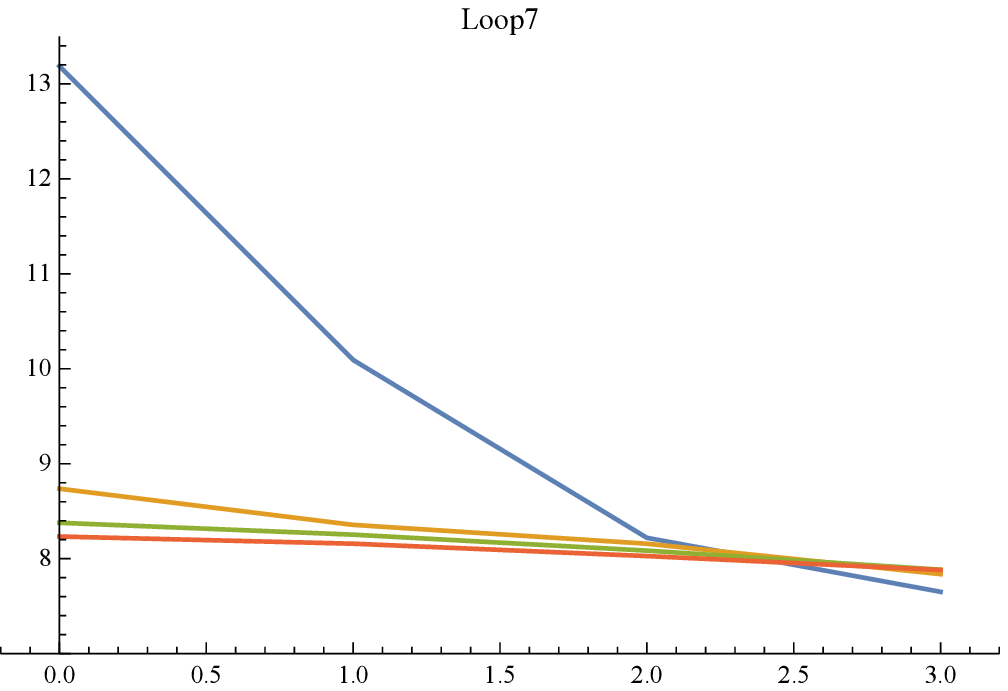} 
\end{center}
\end{minipage}
\begin{minipage}[b]{0.47\linewidth}
\begin{center}
\includegraphics[width=6cm,angle=0,clip]{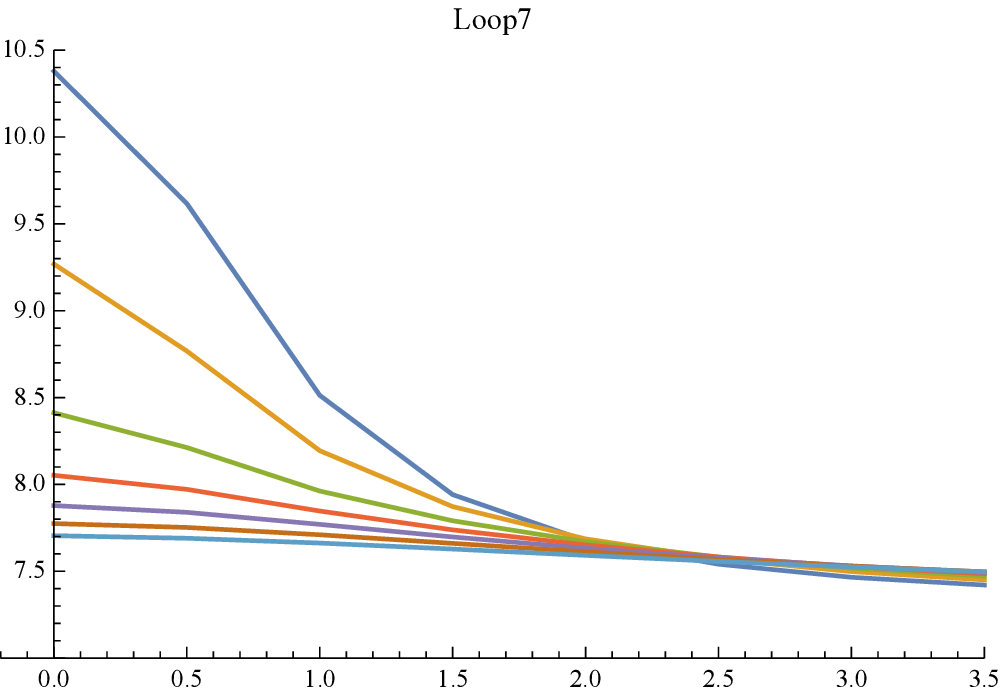} 
\end{center}
\end{minipage}
\caption{ Absolute values of eigenvalues for a fixed $u_1$ in Loop7c (left) and in Loop7d (right). }
\label{L7}
\end{figure*}

The eigenvalues of $Loop 7$ are shown in Fig.\ref{L7}.

\begin{figure*}[htb]
\begin{minipage}[b]{0.47\linewidth}
\begin{center}
\includegraphics[width=6cm,angle=0,clip]{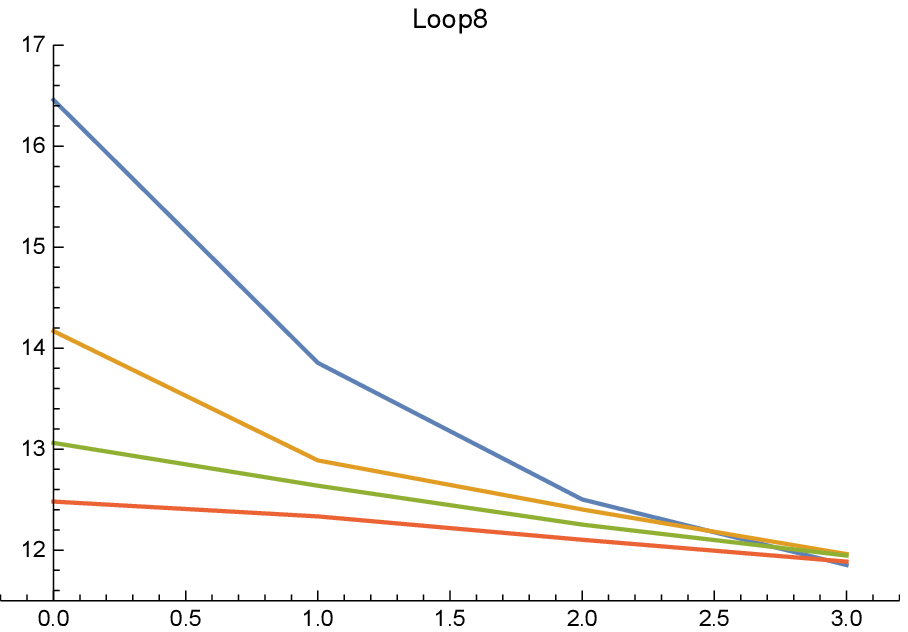} 
\end{center}
\end{minipage}
\begin{minipage}[b]{0.47\linewidth}
\begin{center}
\includegraphics[width=6cm,angle=0,clip]{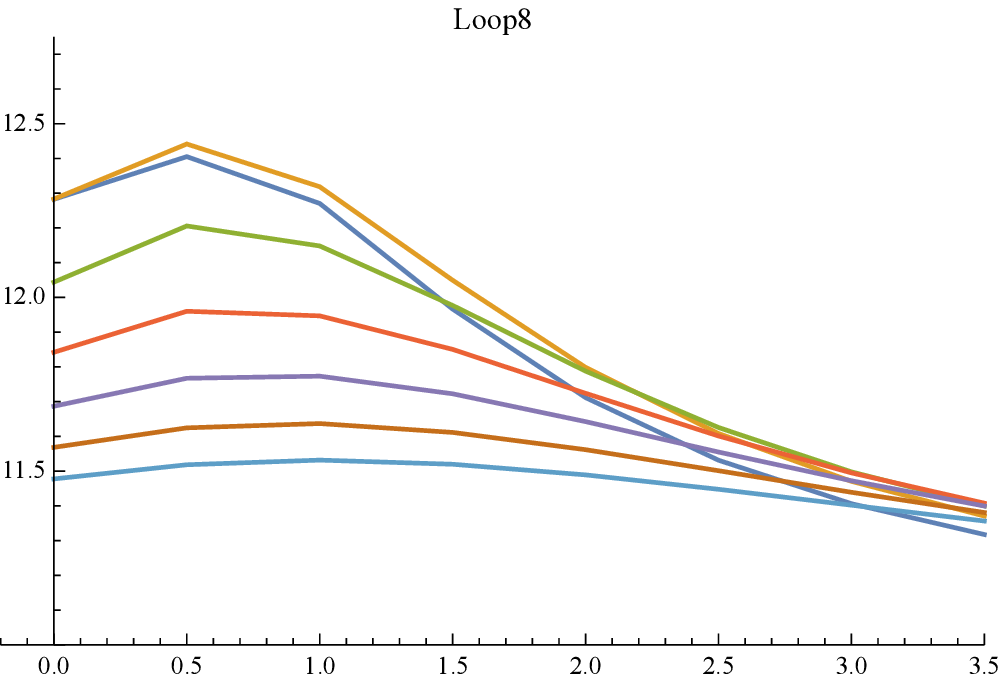} 
\end{center}
\end{minipage}
\caption{ Absolute values of eigenvalues for a fixed $u_1$ in Loop6c (left) and in Loop8d(right).  }
\label{L8}
\end{figure*}

The eigenvalues of $Loop 8$ are shown in Fig.\ref{L8}.

\begin{figure*}[htb]
\begin{minipage}[b]{0.47\linewidth}
\begin{center}
\includegraphics[width=6cm,angle=0,clip]{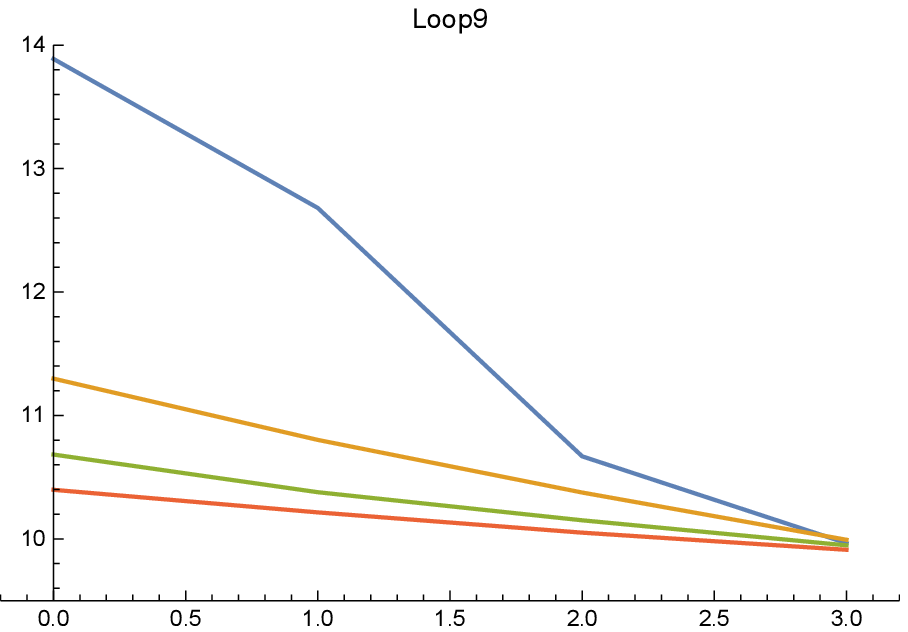}
\end{center}
\end{minipage}
\hfill
\begin{minipage}[b]{0.47\linewidth}
\begin{center}
\includegraphics[width=6cm,angle=0,clip]{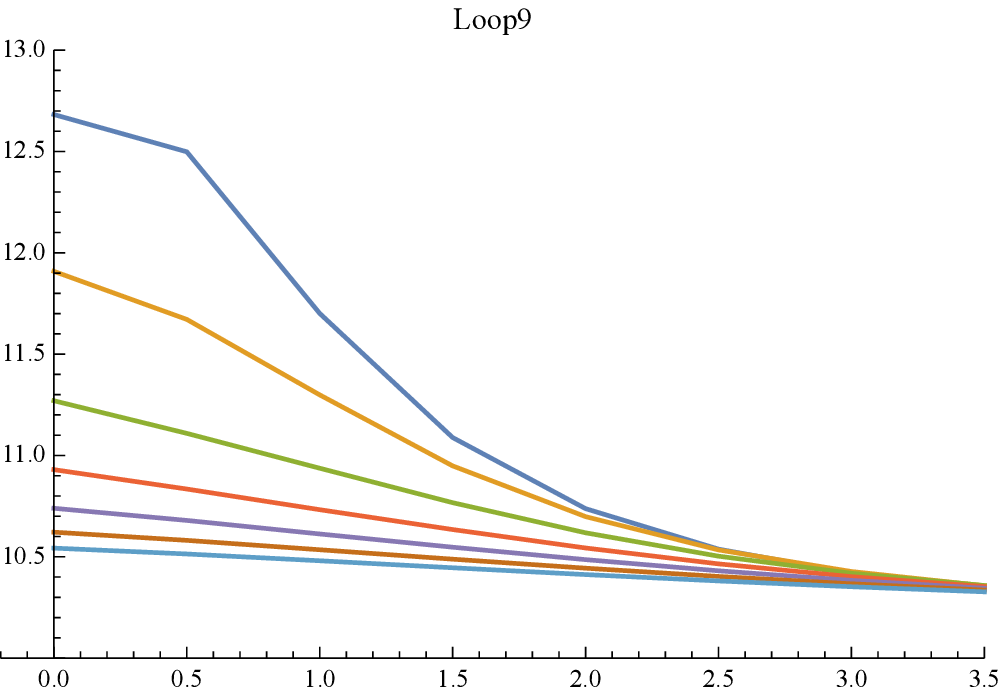}
\end{center}
\end{minipage}
\caption{ Absolute values of eigenvalues for a fixed $u_1$ in Loop9c(left) and in Loop9d(right).  }
\label{L9}
\end{figure*}

The eigenvalues of $Loop 9$ are shown in Fig.\ref{L9} .

\begin{figure*}[htb]
\begin{minipage}[b]{0.47\linewidth}
\begin{center}
\includegraphics[width=6cm,angle=0,clip]{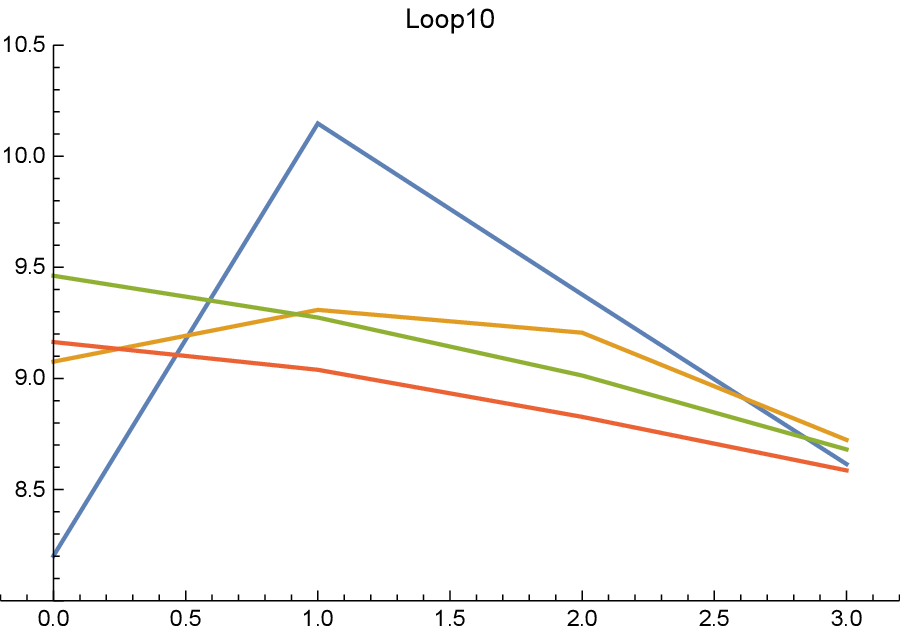} 
\end{center}
\end{minipage}
\begin{minipage}[b]{0.47\linewidth}
\begin{center}
\includegraphics[width=6cm,angle=0,clip]{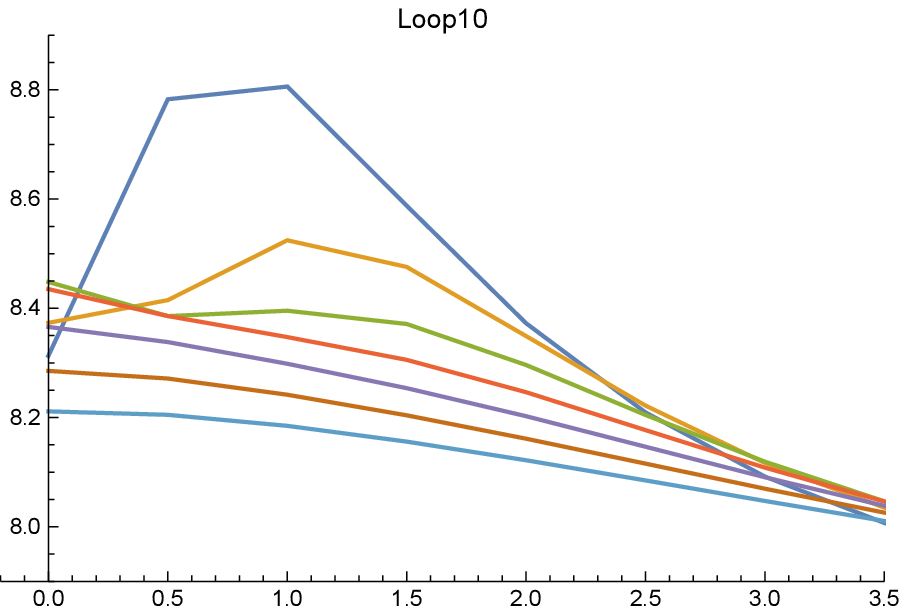} 
\end{center}
\end{minipage}
\caption{ Absolute values of eigenvalues for a fixed $u_1$ in Loop10c(left) and in Loop10d(right). }
\label{L10}
\end{figure*}

\newpage
The eigenvalues of $Loop 10$ are shown in Fig.\ref{L10}.

\begin{figure*}[htb]
\begin{minipage}[b]{0.47\linewidth}
\begin{center}
\includegraphics[width=6cm,angle=0,clip]{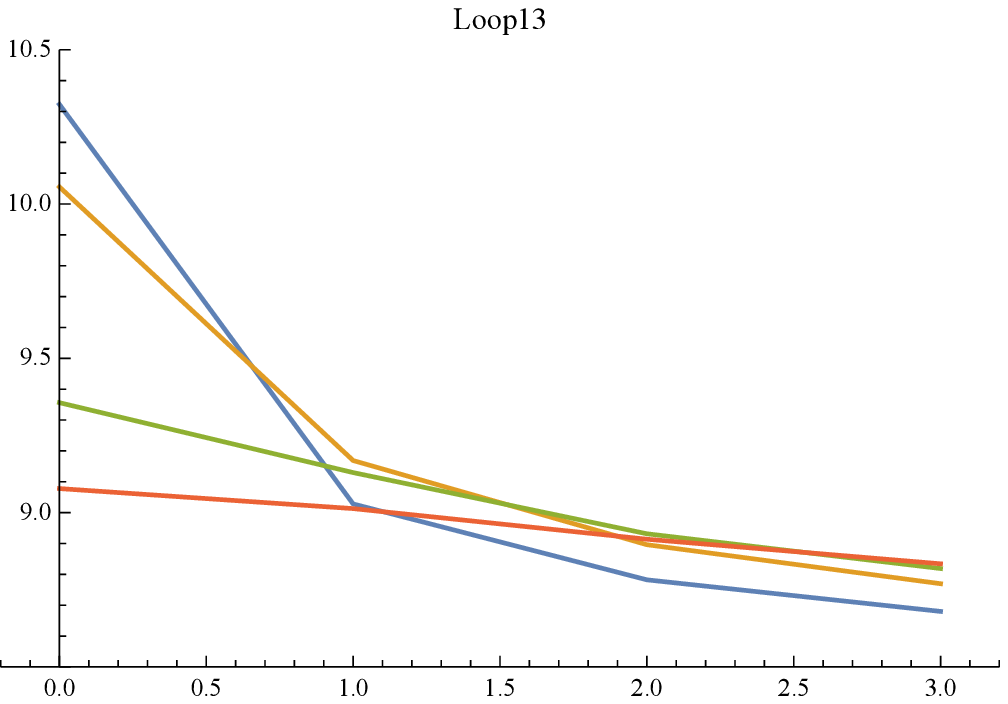}
\end{center}
\end{minipage}
\hfill
\begin{minipage}[b]{0.47\linewidth}
\begin{center}
\includegraphics[width=6cm,angle=0,clip]{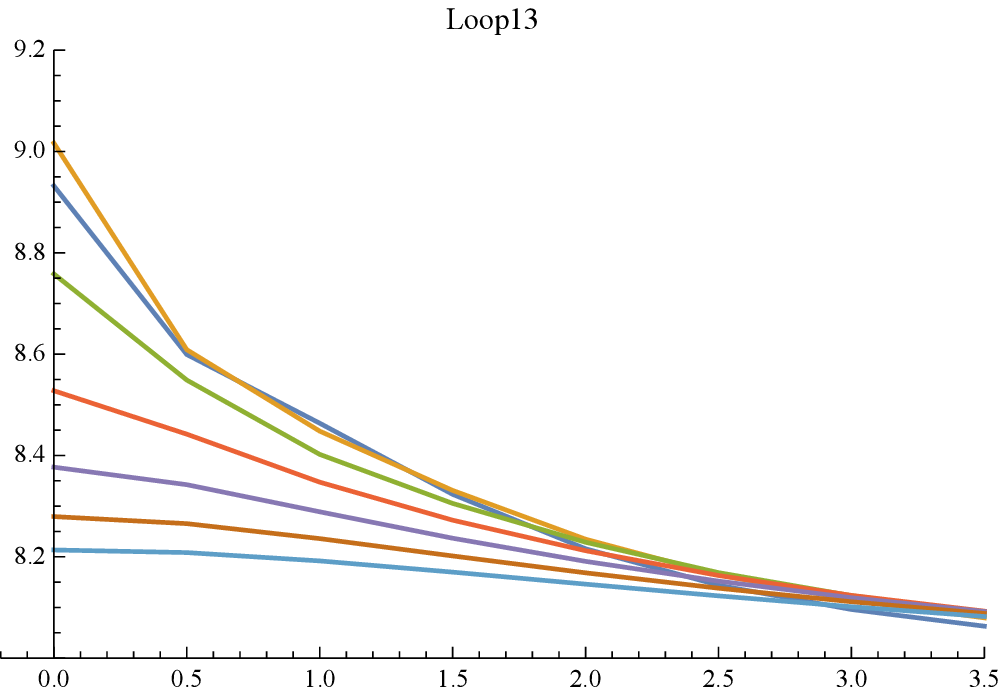}
\end{center}
\end{minipage}
\caption{ Absolute values of eigenvalues for a fixed $u_1$ in Loop13c (left) and in Loop13d(right). }
\label{L13}
\end{figure*}

The eigenvalues of $Loop 13$ are shown in Fig.\ref{L13}.

\begin{figure*}[htb]
\begin{minipage}[b]{0.47\linewidth}
\begin{center}
\includegraphics[width=6cm,angle=0,clip]{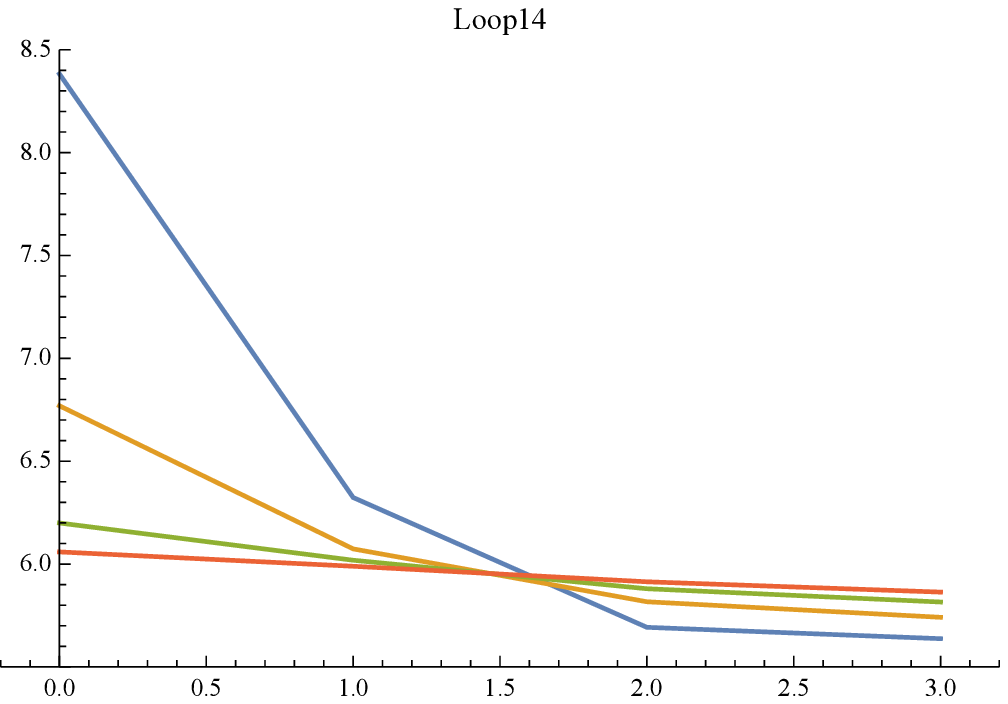}
\end{center}
\end{minipage}
\hfill
\begin{minipage}[b]{0.47\linewidth}
\begin{center}
\includegraphics[width=6cm,angle=0,clip]{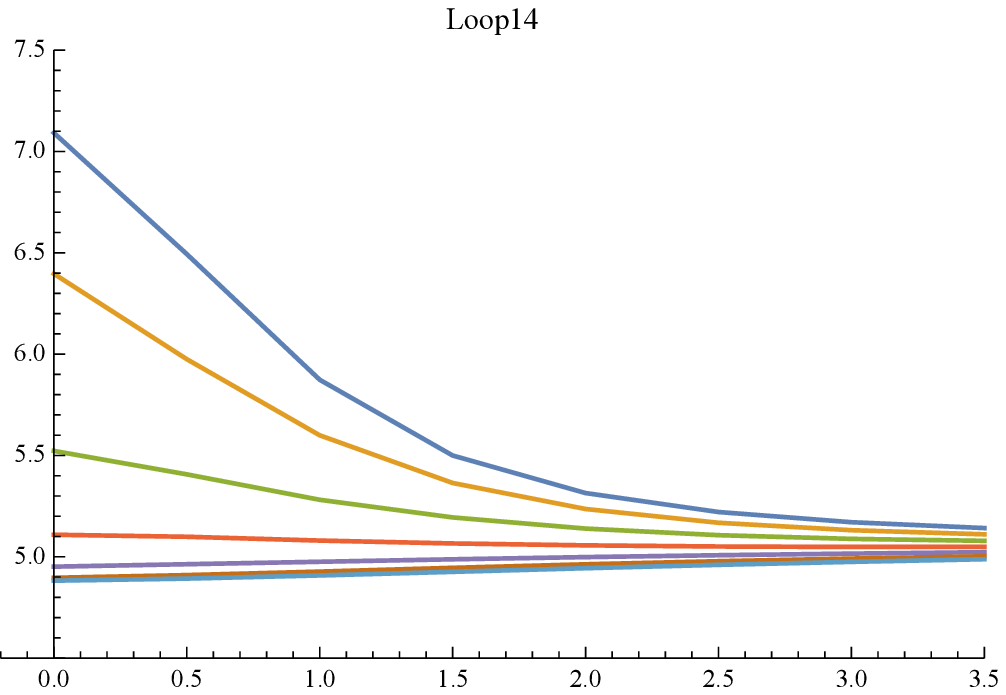}
\end{center}
\end{minipage}
\caption{ Absolute values of eigenvalues for a fixed $u_1$ in Loop14c (left) and in Loop14d(right).  }
\label{L14}
\end{figure*}

The eigenvalues of $Loop 14$ are shown in Fig.\ref{L14} .

\begin{figure*}[htb]
\begin{minipage}[b]{0.47\linewidth}
\begin{center}
\includegraphics[width=6cm,angle=0,clip]{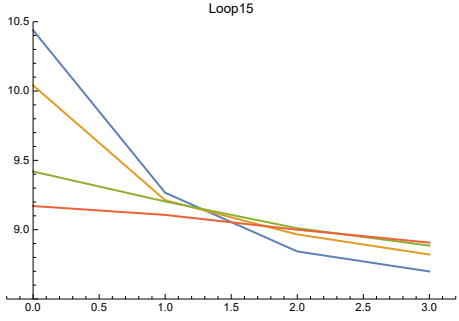}
\end{center}
\end{minipage}
\hfill
\begin{minipage}[b]{0.47\linewidth}
\begin{center}
\includegraphics[width=6cm,angle=0,clip]{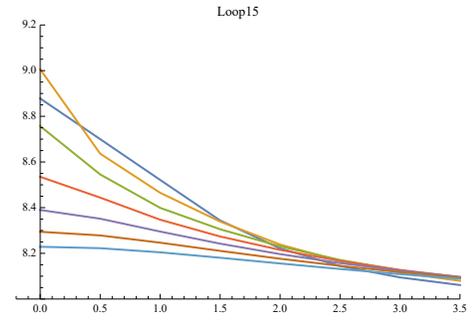}
\end{center}
\end{minipage}
\caption{ Absolute values of eigenvalues for a fixed $u_1$ in Loop15c(left) and in Loop15d (right). }
\label{L15}
\end{figure*}

\begin{figure*}[htb]
\begin{minipage}[b]{0.47\linewidth}
\begin{center}
\includegraphics[width=6cm,angle=0,clip]{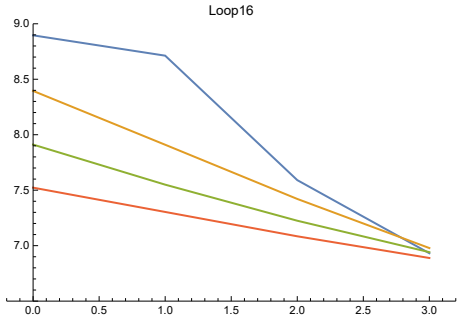}
\end{center}
\end{minipage}
\hfill
\begin{minipage}[b]{0.47\linewidth}
\begin{center}
\includegraphics[width=6cm,angle=0,clip]{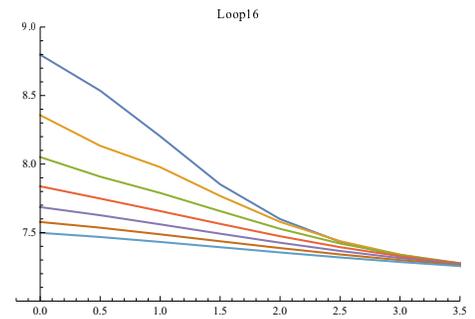}
\end{center}
\end{minipage}
\caption{ Absolute values of eigenvalues for a fixed $u_1$ in Loop16c(left) and in Loop16d(right). }
\label{L16}
\end{figure*}

\newpage
The eigenvalues of $Loop 15$ are shown in Fig.\ref{L15}.
The eigenvalues of $Loop 16$ are shown in Fig.\ref{L16} .

\begin{figure*}[htb]
\begin{minipage}[b]{0.47\linewidth}
\begin{center}
\includegraphics[width=6cm,angle=0,clip]{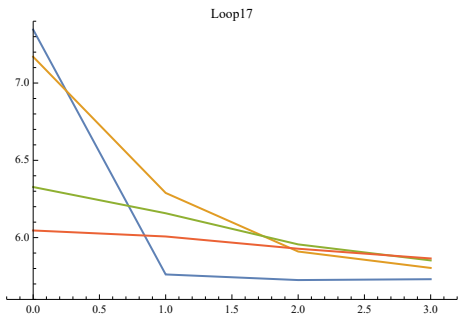}
\end{center}
\end{minipage}
\hfill
\begin{minipage}[b]{0.47\linewidth}
\begin{center}
\includegraphics[width=6cm,angle=0,clip]{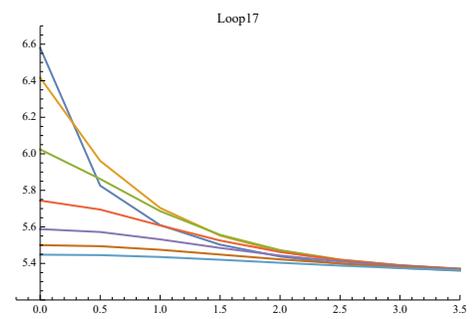}
\end{center}
\end{minipage}
\caption{Absolute values of eigenvalues for a fixed $u_1$ in Loop17c (left) and in Loop17d(right).  }
\label{L17}
\end{figure*}

The eigenvalues of $Loop 17$ are shown in Fig.\ref{L17}.

\begin{figure*}[htb]
\begin{minipage}[b]{0.47\linewidth}
\begin{center}
\includegraphics[width=6cm,angle=0,clip]{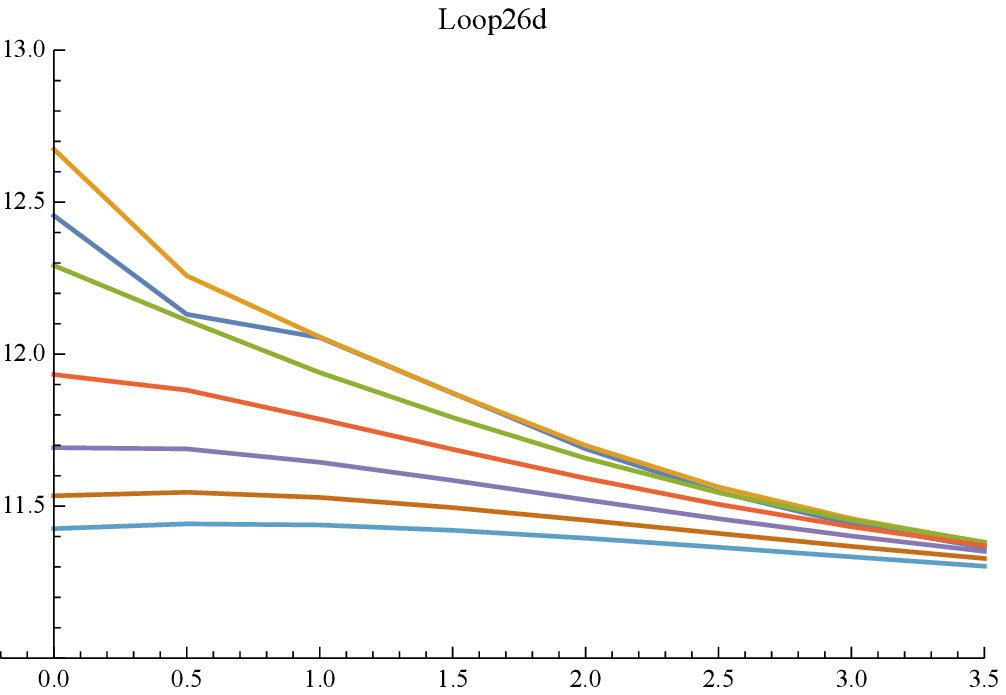} 
\end{center}
\end{minipage}
\hfill
\begin{minipage}[b]{0.47\linewidth}
\begin{center}
\includegraphics[width=6cm,angle=0,clip]{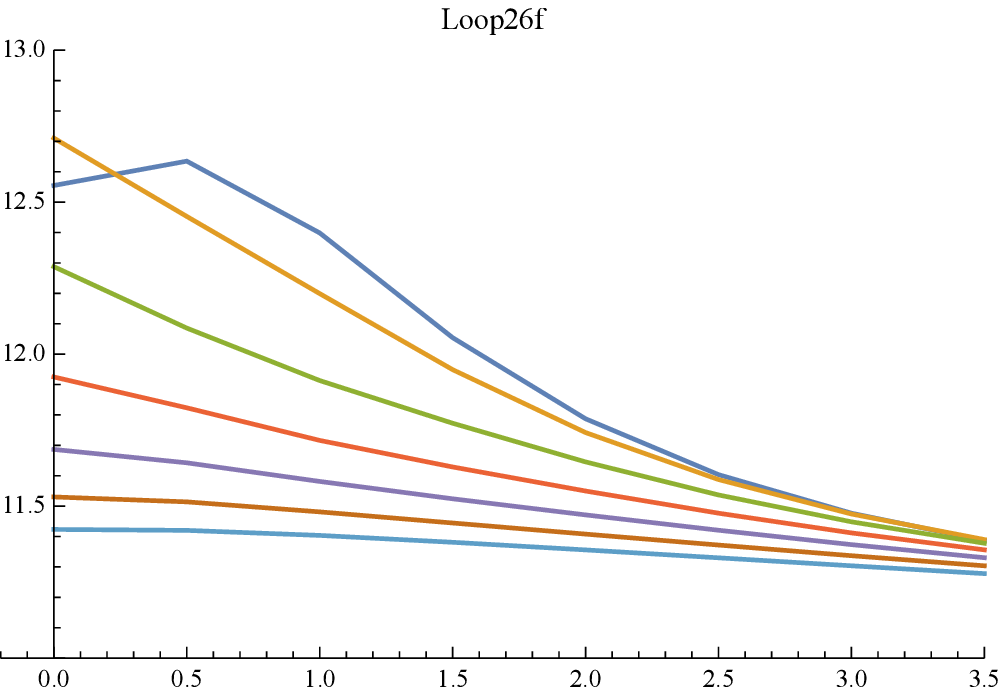} 
\end{center}
\end{minipage}
\caption{ Absolute values of eigenvalues for a fixed $u_1$ in Loop26$\alpha $(left) and in Loop26$\beta$ (right). }
\label{L26}
\end{figure*}

\newpage
The eigenvalues of the $Loop26\alpha$ and $Loop26\beta$ of $a=1/8$ are shown in Fig.\ref{L26}

\begin{figure*}[ht]
\begin{minipage}[b]{0.47\linewidth}
\begin{center}
\includegraphics[width=6cm,angle=0,clip]{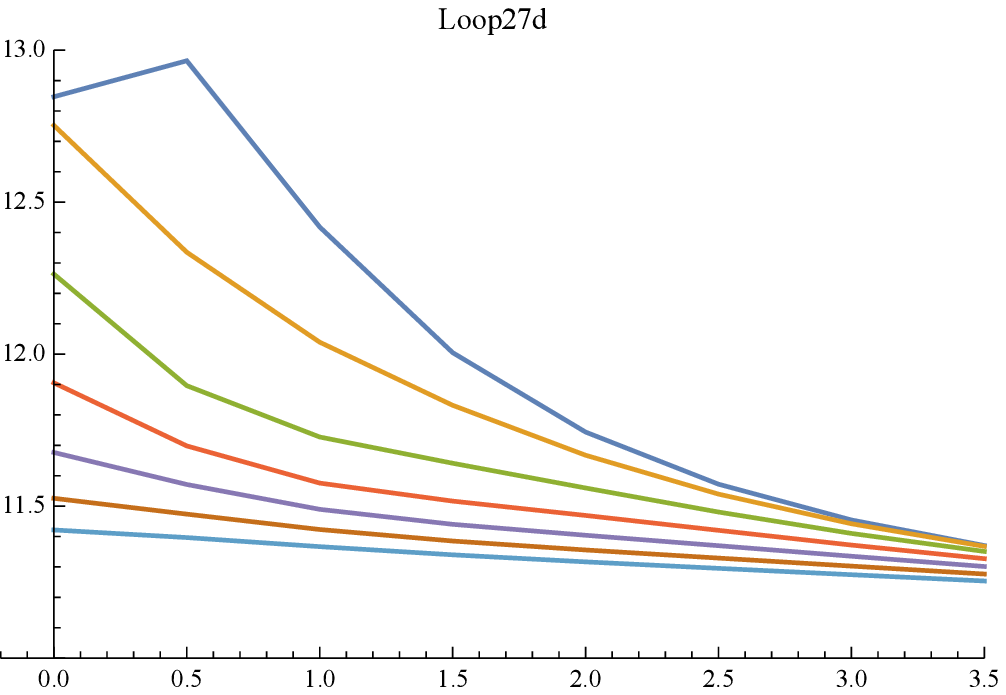} 
\end{center}
\end{minipage}
\hfill
\begin{minipage}[b]{0.47\linewidth}
\begin{center}
\includegraphics[width=6cm,angle=0,clip]{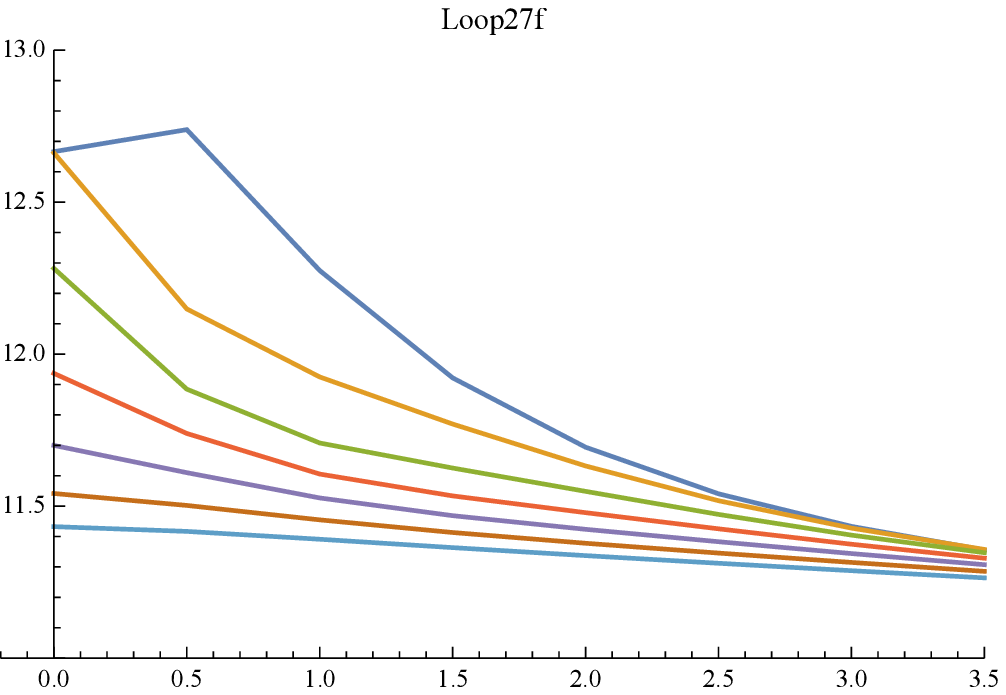} 
\end{center}
\end{minipage}
\caption{ Absolute values of eigenvalues for a fixed $u_1$ in Loop27d (left) and in Loop27f(right). }
\label{L27}
\end{figure*}

The eigenvalues of the $Loop27\alpha$ and $Loop27\beta$ are shown in Fig.\ref{L27}

 \section{Lattice spacing dependence of traces of link variables}
 In Clifford algebra, transformation of a coordinate $X$ is represented by\cite{Porteous95}
 \[
 \left(\begin{array}{cc}
       a& c\\
       b & d\end{array}\right)\left(\begin{array}{cc}
               x& x x^-\\
               1& x^-\end{array}\right)\left(\begin{array}{cc}
                d^-& c^-\\
                 b^-&a^-\end{array}\right)=\lambda\left(\begin{array}{cc}
                    x'& x'{x'}^-\\
                    1&{x}'^-\end{array}\right)
 \]
 and the term $x x^-$ yields link products.
 
In lattice simulations of scalar fields, we consider Feynman path integrals is $Z=\int[d\psi]e^{-S}$ with ${\bf x}=a (u_1 e_1+u_2 e_2)+a u_3 e_1\wedge e_2$, 
$-N/2<u_1,u_2,u_3 \leq N/2$, $\mu=1,2,3$. $N\sim 2^{11}=2048$. The scale of $e_1\wedge e_2$ is chosen to be the same as $e_1,e_2$, for Wilson loops, but it can be complex for Polyakov loops.

The expectation values of Wilson or Polyakov action $S$ consists of eigenvalues of  left lower components of the Loop matrices $Lk[u_1,u_2]$ where $k$ specifies the FP actions of \cite{DGHHN95}, and the trace of $D{\bf S}(Lk[u_1,u_2])$ which consists of the sum of $dS_1$ and $dS_2$ along the loops

The $4\times 4$ matrix of the $Loop1$ contribution
\[
D{\bf S}(L1[u_1,u_2])=dS_1[u_1,u_2]+dS_2[u_1+a,u_2]-dS_1[u_1+a,u_2+a]-dS_2[u_1,u_2+a]
\]
has non-zero components in the right upper component. We measure the trace of the $2\times 2$ matrices.

In the case of $Loop 2$,
\[
D{\bf S}(L2[u_1,u_2])=dS_1[u_1,u_2]+2 dS_2[u_1+a,u_2]-dS_1[u_1+a,u_2+2a]-2 dS_2[u_1,u_2+2a]
\]
has non-zero components only in the right upper corner.
\newpage
\subsection{Paths on one 2$D$ plane expanded by $e_1$ and $e_2$}

\begin{figure*}[htb]
\begin{minipage}[b]{0.47\linewidth}
\begin{center}
\includegraphics[width=6cm,angle=0,clip]{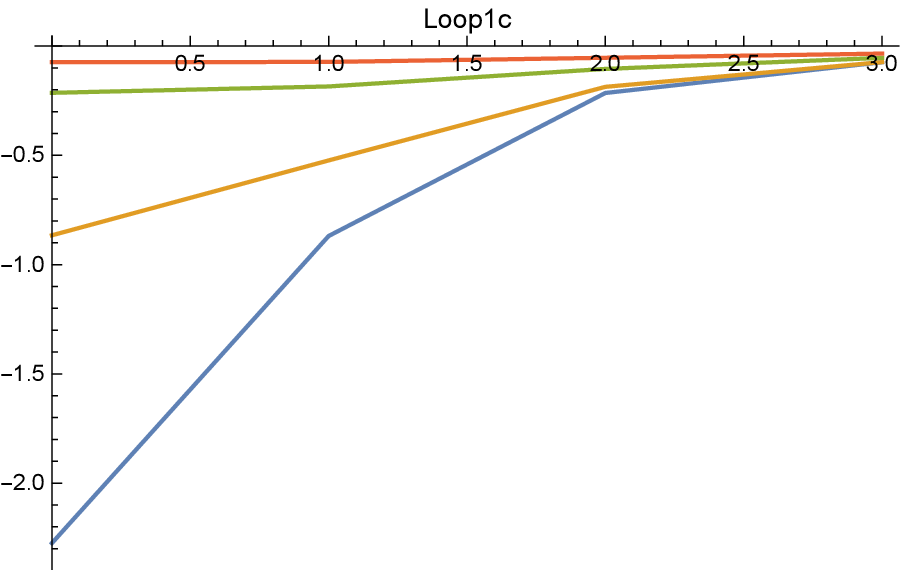}
\end{center}
\end{minipage}
\hfill
\begin{minipage}[b]{0.47\linewidth}
\begin{center}
\includegraphics[width=6cm,angle=0,clip]{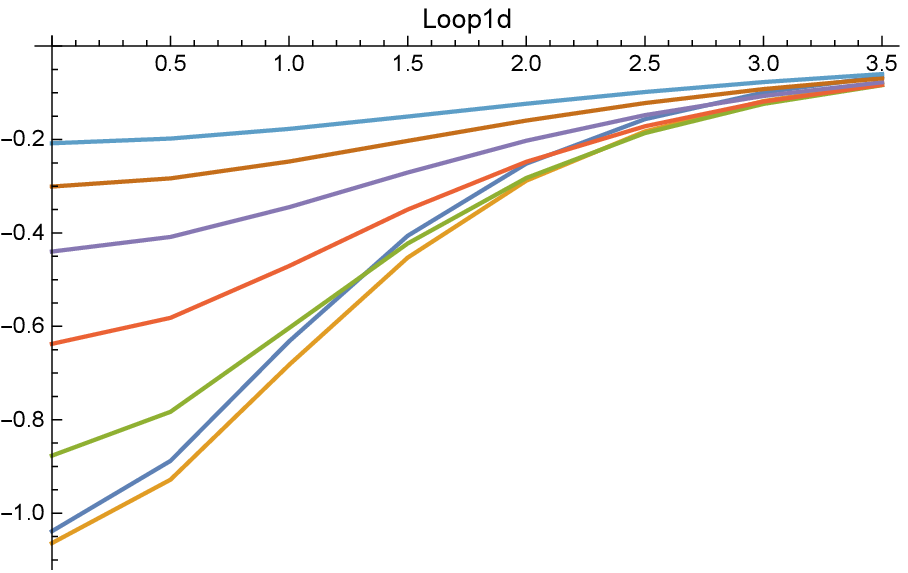}
\end{center}
\end{minipage}
\caption{Trace of $d{\bf S}(L1[u_1,u_2])$ for $\Delta{u_i}=1 $(left) and $\Delta{u_i}=\frac{1}{2} $(right). ($i=1,2$) }
\label{L1tr}
\end{figure*}

The traces of the $2\times 2$ matrix, which is twice the real part of the diagonal component in the case of $Loop1$ as a function of $u_2$ for fixed $u_1$ are shown in Fig.\ref{L1tr}.

\begin{figure*}[htb]
\begin{minipage}[b]{0.47\linewidth}
\begin{center}
\includegraphics[width=6cm,angle=0,clip]{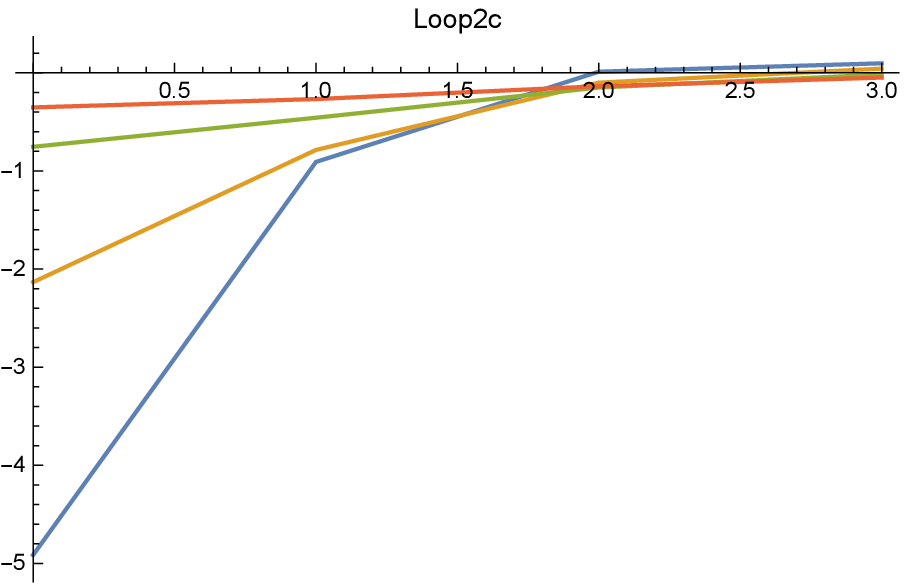}
\end{center}
\end{minipage}
\hfill
\begin{minipage}[b]{0.47\linewidth}
\begin{center}
\includegraphics[width=6cm,angle=0,clip]{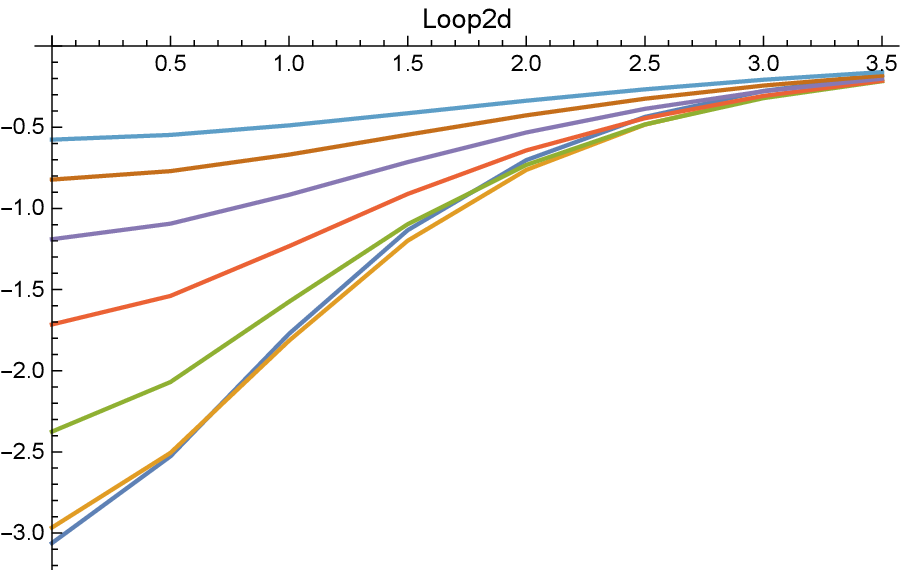}
\end{center}
\end{minipage}
\caption{Trace of $d{\bf S}(L2[u_1,u_2])$ for $\Delta{u_i}=1$(left) and $\Delta{u_i}=\frac{1}{2}$(right). ($i=1,2)$ }
\label{L2tr}
\end{figure*}

The traces of the matrix in the case of $Loop2$ are shown in Fig.\ref{L2tr}.

\begin{figure*}[htb]
\begin{minipage}[b]{0.47\linewidth}
\begin{center}
\includegraphics[width=6cm,angle=0,clip]{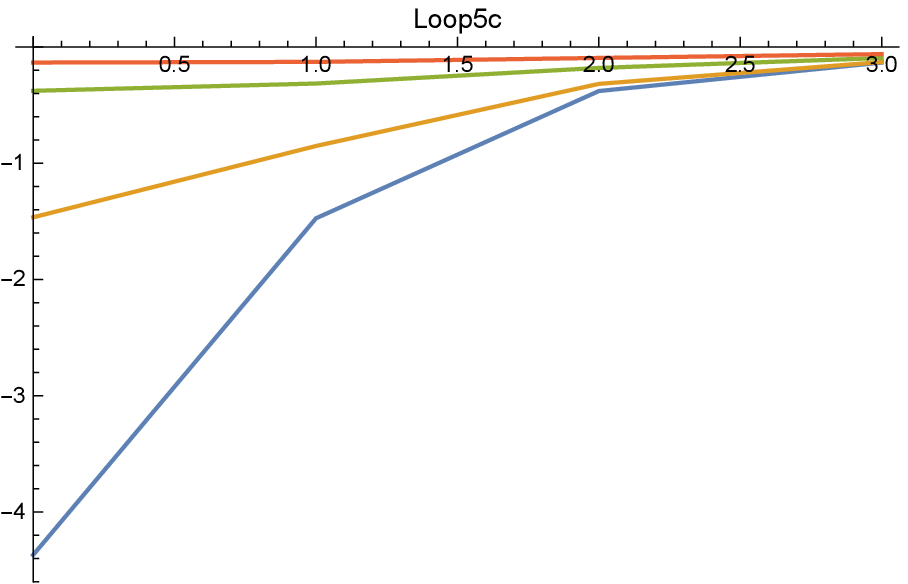}
\end{center}
\end{minipage}
\hfill
\begin{minipage}[b]{0.47\linewidth}
\begin{center}
\includegraphics[width=6cm,angle=0,clip]{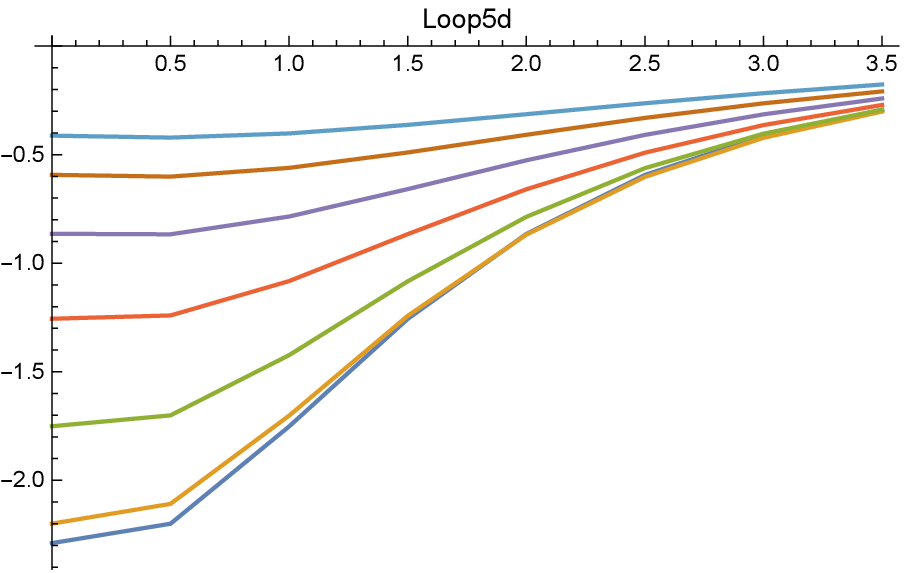}
\end{center}
\end{minipage}
\caption{Trace of $d{\bf S}(L5[u_1,u_2])$ for $\Delta{u_i}=1$(left) and $\Delta{u_i}=\frac{1}{2}$(right). ($i=1,2)$ }
\label{L5tr}
\end{figure*}

The traces of the matrix in the case of $Loop5$ are shown in Fig.\ref{L5tr}.

\begin{figure*}[htb]
\begin{minipage}[b]{0.47\linewidth}
\begin{center}
\includegraphics[width=6cm,angle=0,clip]{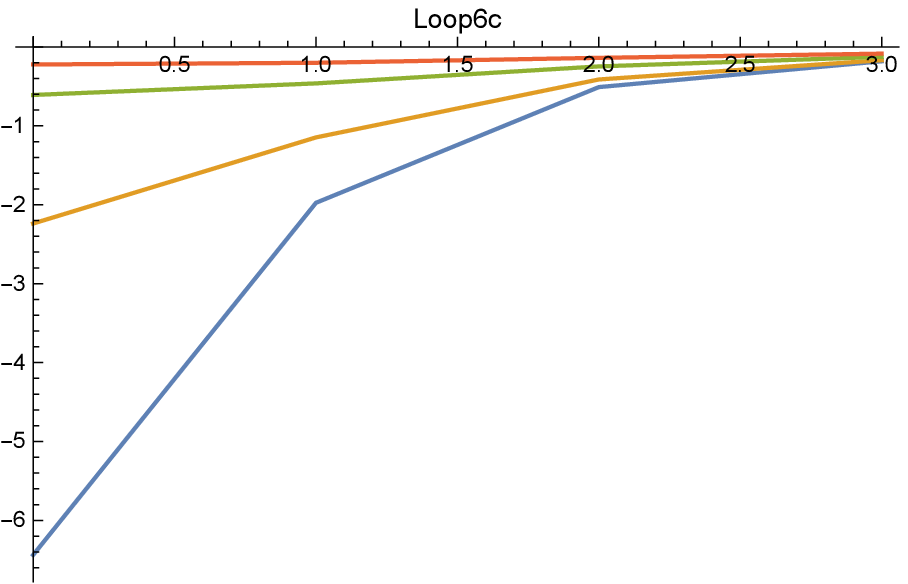}
\end{center}
\end{minipage}
\hfill
\begin{minipage}[b]{0.47\linewidth}
\begin{center}
\includegraphics[width=6cm,angle=0,clip]{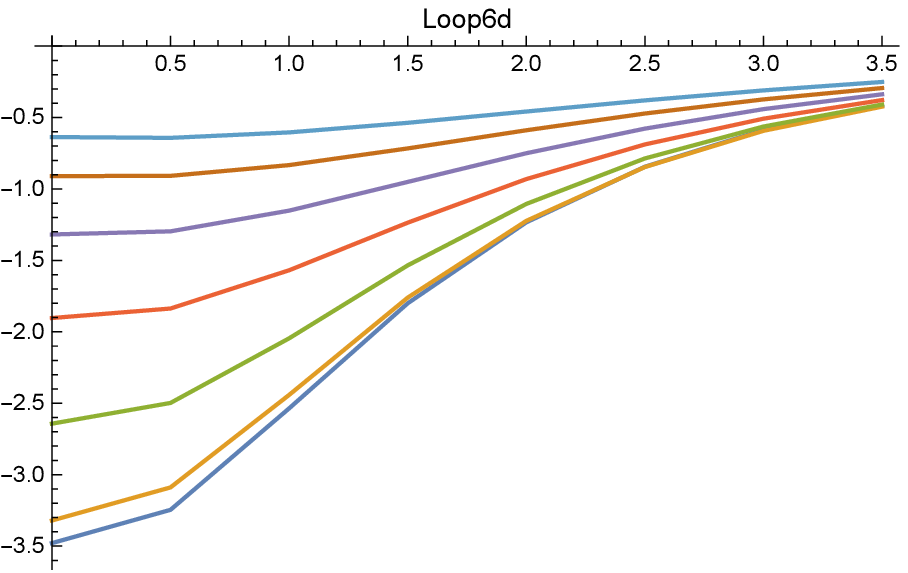}
\end{center}
\end{minipage}
\caption{Trace of $d{\bf S}(L6[u_1,u_2])$ for $\Delta{u_i}=1$(left) and $\Delta{u_i}=\frac{1}{2}$(right). ($i=1,2)$ }
\label{L6tr}
\end{figure*}

\newpage
The traces of the matrix in the case of $Loop5$ are shown in  Fig.\ref{L6tr}.

\begin{figure*}[htb]
\begin{minipage}[b]{0.47\linewidth}
\begin{center}
\includegraphics[width=6cm,angle=0,clip]{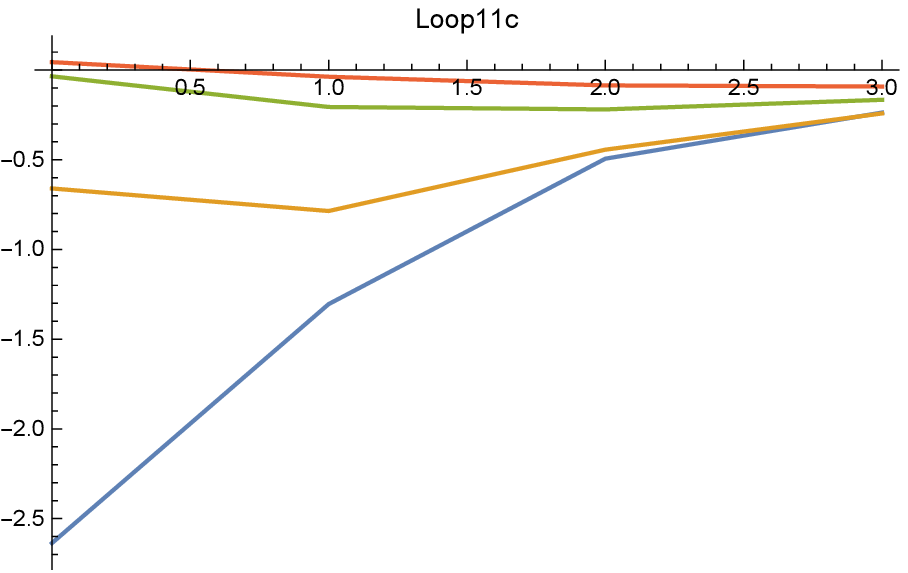}
\end{center}
\end{minipage}
\hfill
\begin{minipage}[b]{0.47\linewidth}
\begin{center}
\includegraphics[width=6cm,angle=0,clip]{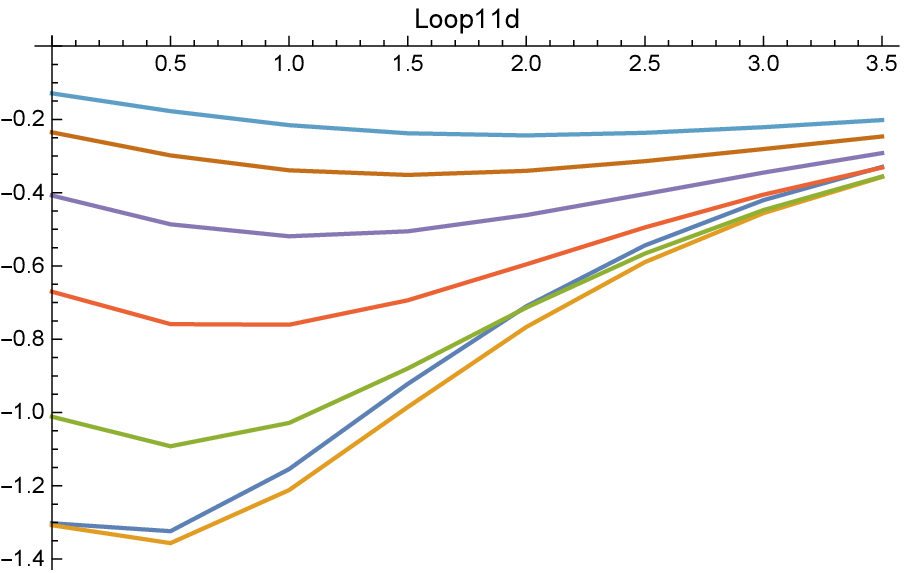}
\end{center}
\end{minipage}
\caption{Trace of $d{\bf S}(L11[u_1,u_2])$ for $\Delta{u_i}=1$(left) and $\Delta{u_i}=\frac{1}{2}$(right). ($i=1,2)$ }
\label{L11tr}
\end{figure*}

The traces of the matrix in the case of $Loop11$ are shown in  Fig.\ref{L11tr}.

\begin{figure*}[htb]
\begin{minipage}[b]{0.47\linewidth}
\begin{center}
\includegraphics[width=6cm,angle=0,clip]{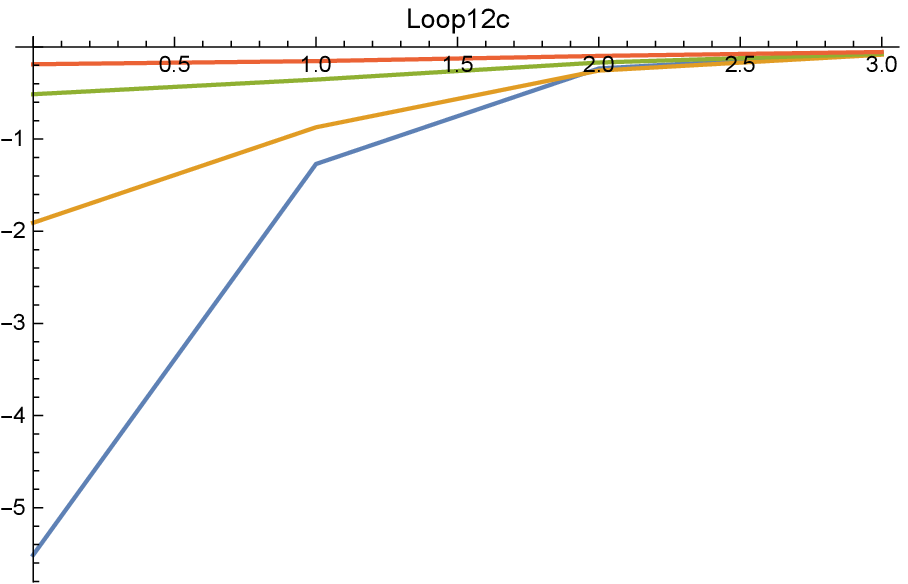}
\end{center}
\end{minipage}
\hfill
\begin{minipage}[b]{0.47\linewidth}
\begin{center}
\includegraphics[width=6cm,angle=0,clip]{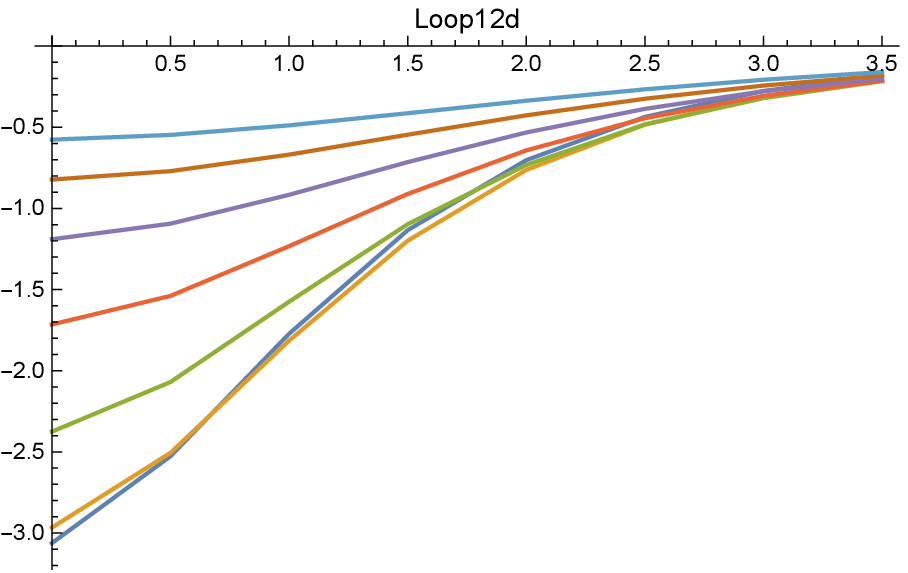}
\end{center}
\end{minipage}
\caption{Trace of $d{\bf S}(L12[u_1,u_2])$ for $\Delta{u_i}=1 $(left) and $\Delta{u_i}=\frac{1}{2} $(right). ($i=1,2)$ }
\label{L12tr}
\end{figure*}

The traces of the matrix in the case of $Loop12$ are shown in  Fig.\ref{L12tr}.

\begin{figure*}[htb]
\begin{minipage}[b]{0.47\linewidth}
\begin{center}
\includegraphics[width=6cm,angle=0,clip]{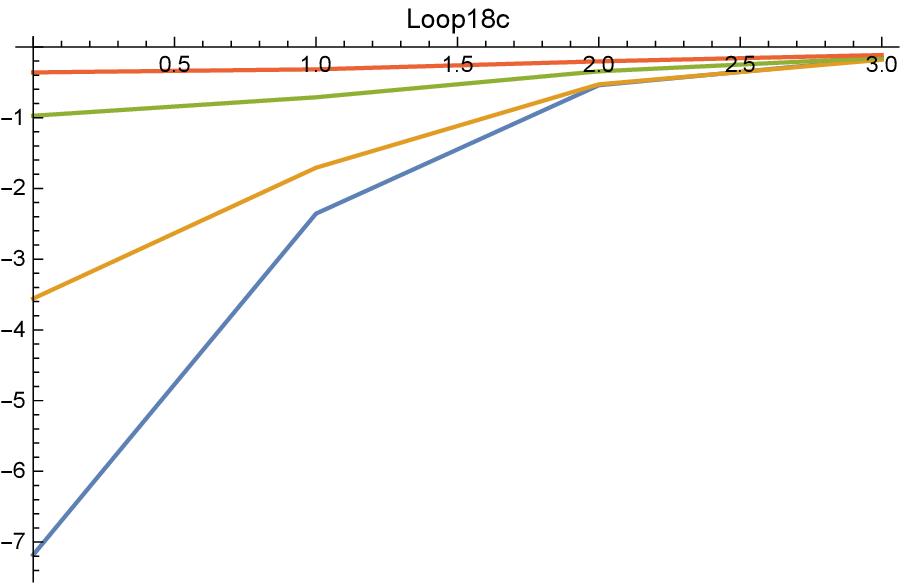}
\end{center}
\end{minipage}
\hfill
\begin{minipage}[b]{0.47\linewidth}
\begin{center}
\includegraphics[width=6cm,angle=0,clip]{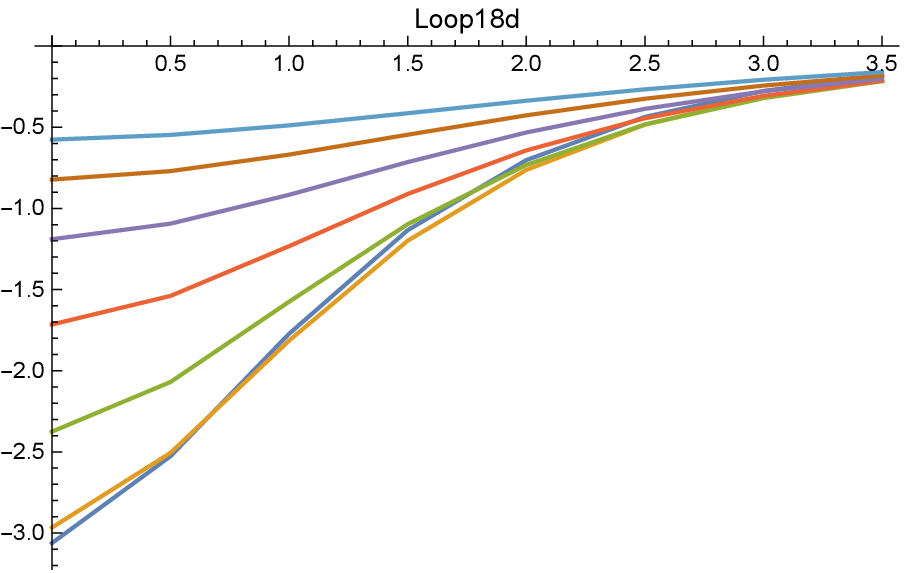}
\end{center}
\end{minipage}
\caption{Trace of $d{\bf S}(L18[u_1,u_2])$ for $\Delta{u_i}=1$(left) and $\Delta{u_i}=\frac{1}{2} $(right). ($i=1,2)$ }
\label{L18tr}
\end{figure*}

\newpage
The traces of the matrix in the case of $Loop18$ are shown in  Fig.\ref{L18tr}..

\subsection{Paths on two planes connected by $e_1\wedge e_2$}

\begin{figure*}[htb]
\begin{minipage}[b]{0.47\linewidth}
\begin{center}
\includegraphics[width=6cm,angle=0,clip]{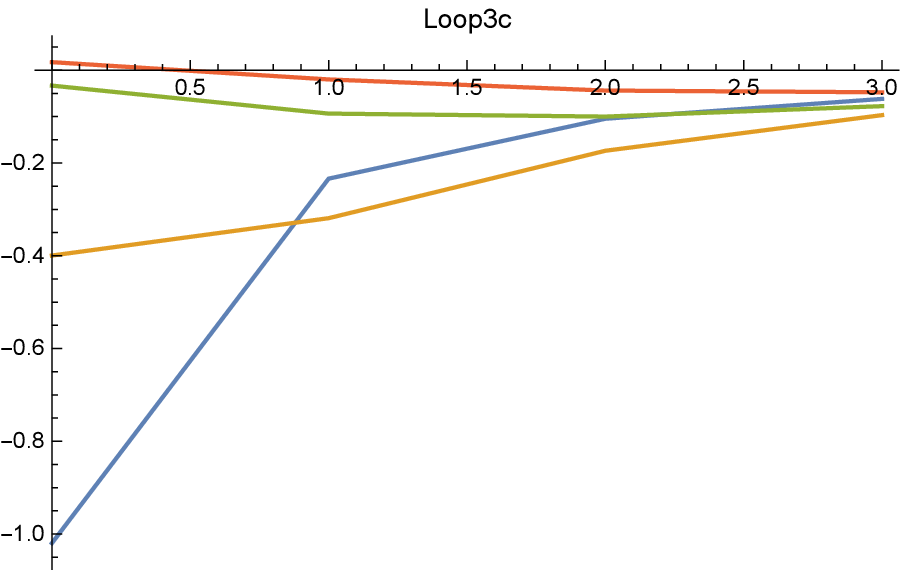}
\end{center}
\end{minipage}
\hfill
\begin{minipage}[b]{0.47\linewidth}
\begin{center}
\includegraphics[width=6cm,angle=0,clip]{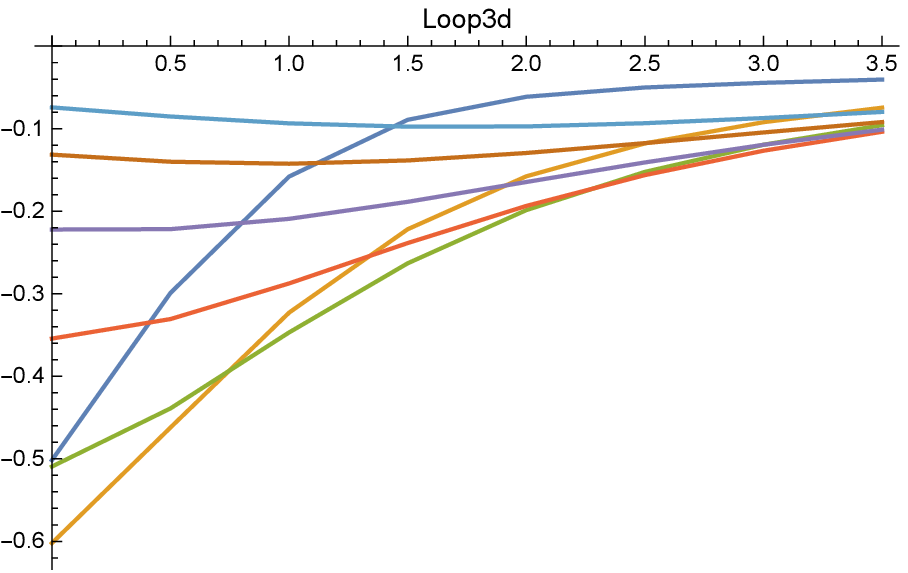}
\end{center}
\end{minipage}
\caption{Trace of $d{\bf S}(L3[u_1,u_2])$ for $\Delta{u_i}=1 $(left) and $\Delta{u_i}=\frac{1}{2} $(right). ($i=1,2)$ }
\label{L3tr}
\end{figure*}

The traces of the matrix in the case of $Loop3$ are shown in Fig.\ref{L3tr}.

\begin{figure*}[htb]
\begin{minipage}[b]{0.47\linewidth}
\begin{center}
\includegraphics[width=6cm,angle=0,clip]{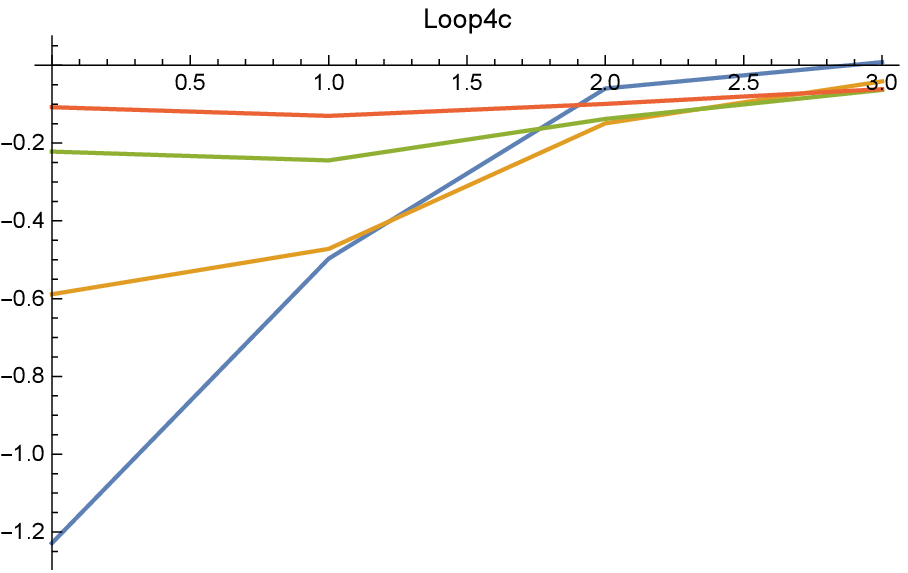}
\end{center}
\end{minipage}
\hfill
\begin{minipage}[b]{0.47\linewidth}
\begin{center}
\includegraphics[width=6cm,angle=0,clip]{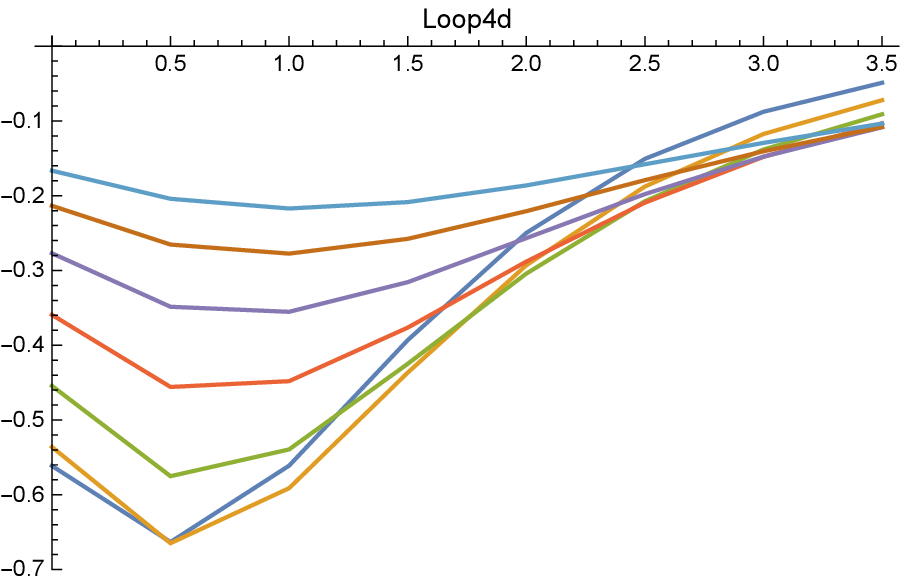}
\end{center}
\end{minipage}
\caption{Trace of $d{\bf S}(L4[u_1,u_2])$ for $\Delta{u_i}=1 $(left) and $\Delta{u_i}=\frac{1}{2} $(right). ($i=1,2)$ }
\label{L4tr}
\end{figure*}

The traces of the matrix in the case of $Loop4$ are shown in  Fig.\ref{L4tr}.

\begin{figure*}[htb]
\begin{minipage}[b]{0.47\linewidth}
\begin{center}
\includegraphics[width=6cm,angle=0,clip]{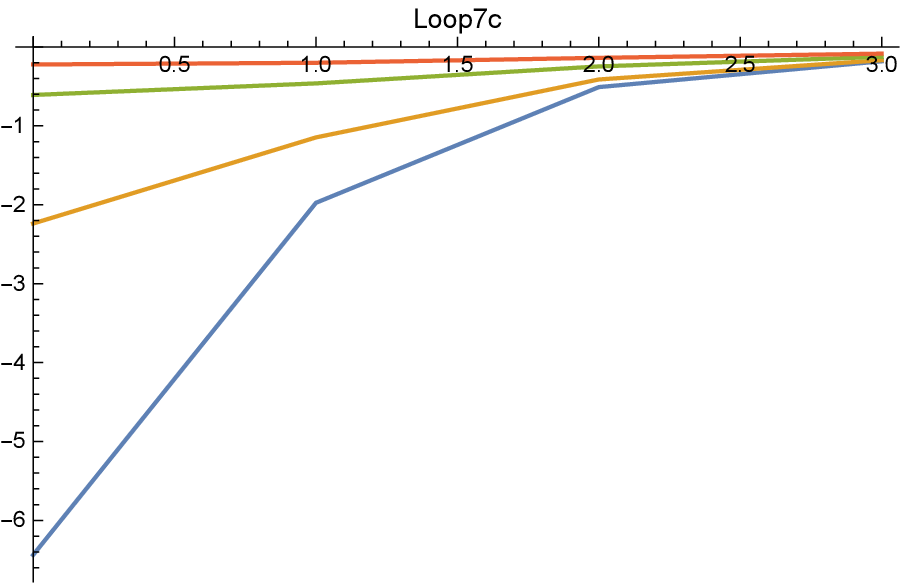}
\end{center}
\end{minipage}
\hfill
\begin{minipage}[b]{0.47\linewidth}
\begin{center}
\includegraphics[width=6cm,angle=0,clip]{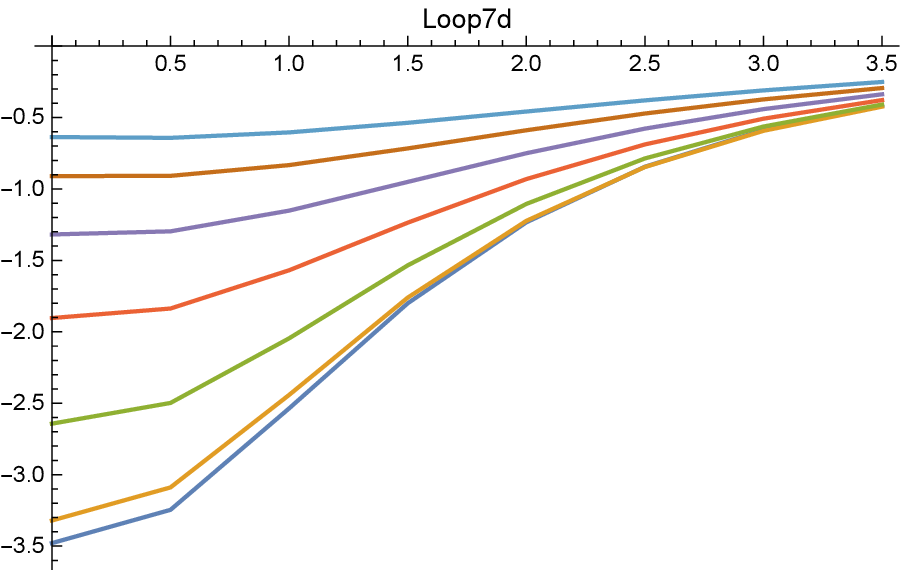}
\end{center}
\end{minipage}
\caption{Trace of $d{\bf S}(L7[u_1,u_2])$ for $\Delta{u_i}=1 $(left) and $\Delta{u_i}=\frac{1}{2} $ (right). ($i=1,2)$ }
\label{L7tr}
\end{figure*}

\newpage
The traces of the matrix in the case of $Loop7$ are shown in Fig.\ref{L7tr}.

\begin{figure*}[htb]
\begin{minipage}[b]{0.47\linewidth}
\begin{center}
\includegraphics[width=6cm,angle=0,clip]{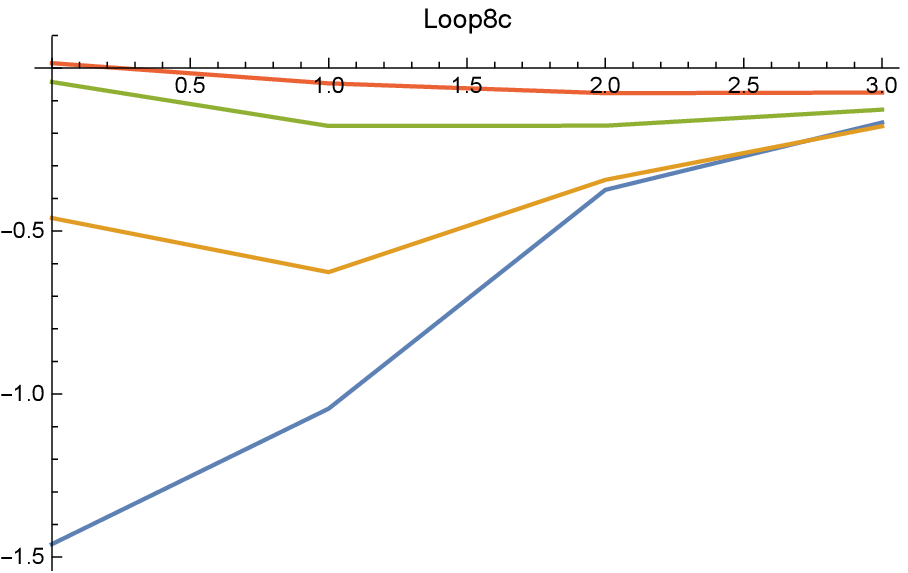}
\end{center}
\end{minipage}
\hfill
\begin{minipage}[b]{0.47\linewidth}
\begin{center}
\includegraphics[width=6cm,angle=0,clip]{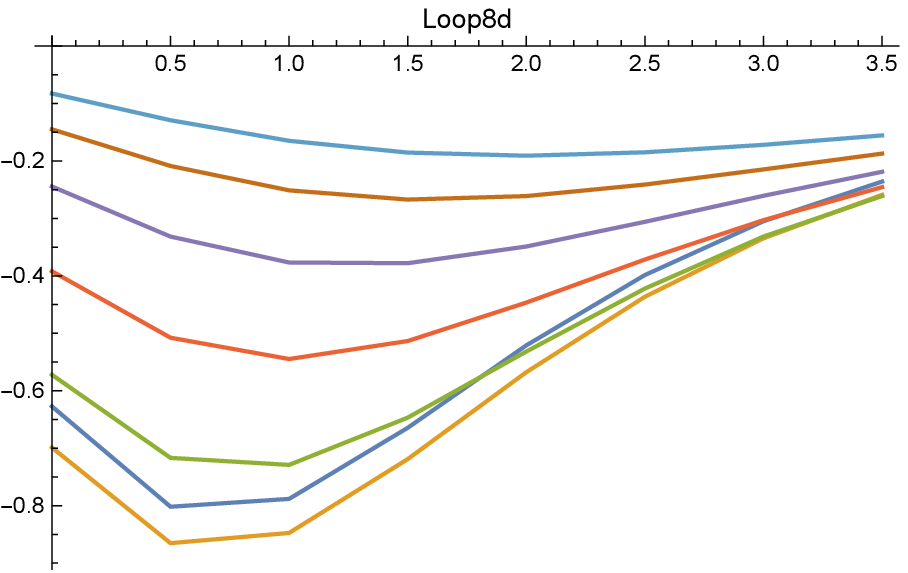}
\end{center}
\end{minipage}
\caption{Trace of $d{\bf S}(L8[u_1,u_2])$ for $\Delta{u_i}=1 $(left) and $\Delta{u_i}=\frac{1}{2} $ (right.) ($i=1,2)$ }
\label{L8tr}
\end{figure*}

The traces of the matrix in the case of $Loop8$ are shown in Fig.\ref{L8tr}.

\begin{figure*}[htb]
\begin{minipage}[b]{0.47\linewidth}
\begin{center}
\includegraphics[width=6cm,angle=0,clip]{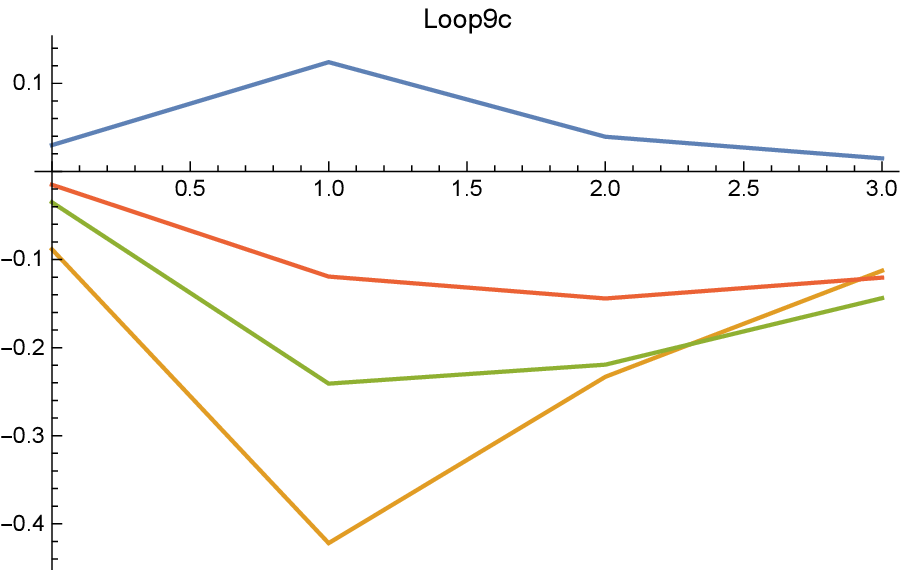}
\end{center}
\end{minipage}
\hfill
\begin{minipage}[b]{0.47\linewidth}
\begin{center}
\includegraphics[width=6cm,angle=0,clip]{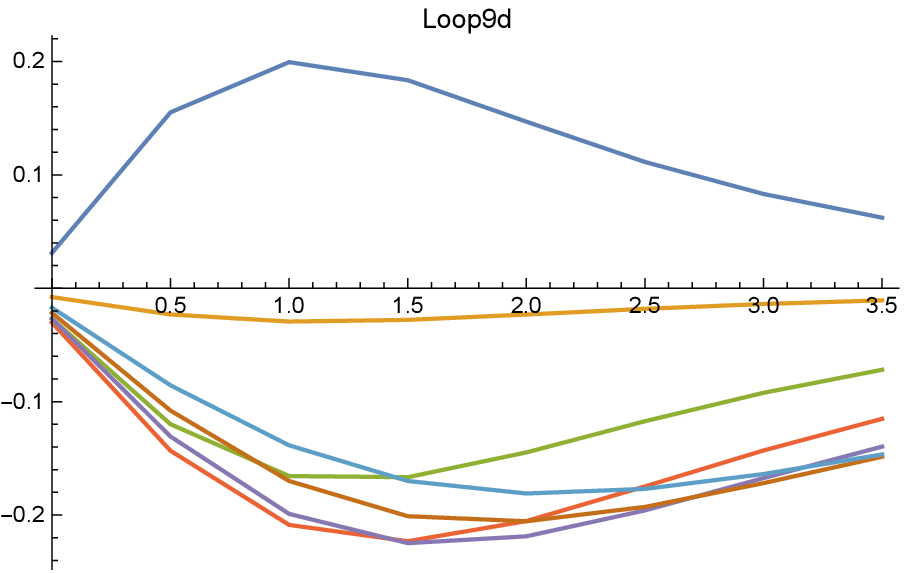}
\end{center}
\end{minipage}
\caption{Trace of $d{\bf S}(L9[u_1,u_2])$ for $\Delta{u_i}=1 $(left) and $\Delta{u_i}=\frac{1}{2} $ (right). ($i=1,2)$ }
\label{L9tr}
\end{figure*}

The traces of the matrix in the case of $Loop9$ are shown in Fig.\ref{L9tr}.

\begin{figure*}[htb]
\begin{minipage}[b]{0.47\linewidth}
\begin{center}
\includegraphics[width=6cm,angle=0,clip]{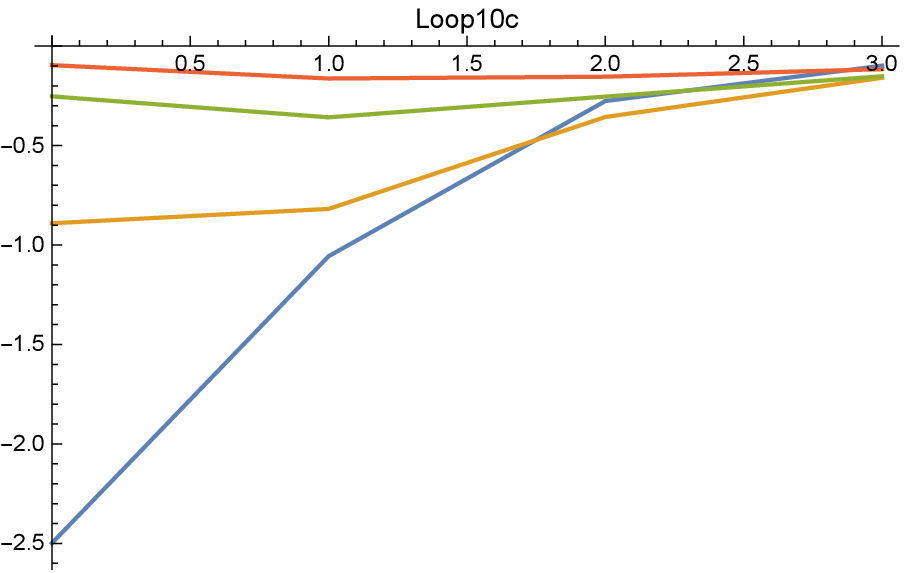}
\end{center}
\end{minipage}
\hfill
\begin{minipage}[b]{0.47\linewidth}
\begin{center}
\includegraphics[width=6cm,angle=0,clip]{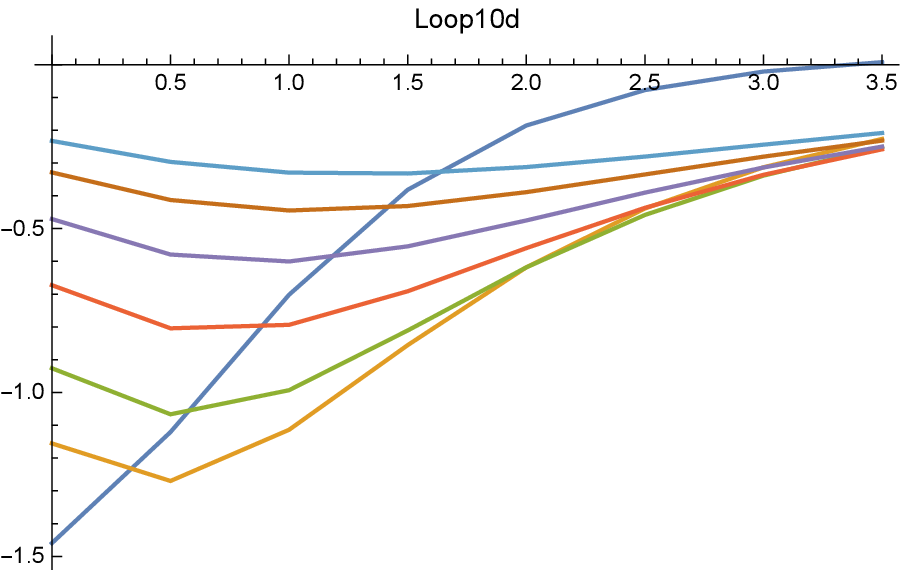}
\end{center}
\end{minipage}
\caption{Trace of $d{\bf S}(L10[u_1,u_2])$ for $\Delta{u_i}=1 $(left) and $\Delta{u_i}=\frac{1}{2} $ (right). ($i=1,2)$ }
\label{L10tr}
\end{figure*}

\newpage
The traces of the matrix in the case of $Loop10$ are shown in Fig.\ref{L10tr}.

\begin{figure*}[htb]
\begin{minipage}[b]{0.47\linewidth}
\begin{center}
\includegraphics[width=6cm,angle=0,clip]{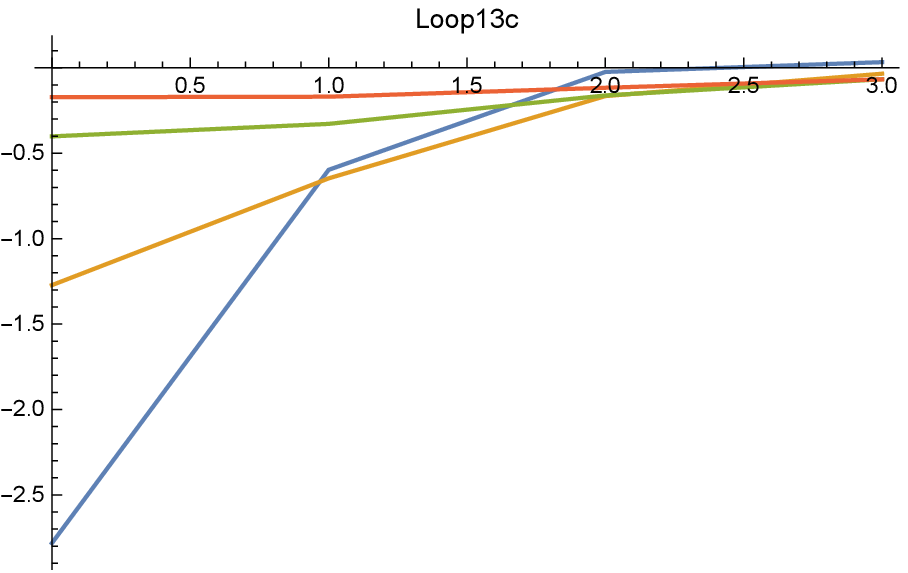}
\end{center}
\end{minipage}
\hfill
\begin{minipage}[b]{0.47\linewidth}
\begin{center}
\includegraphics[width=6cm,angle=0,clip]{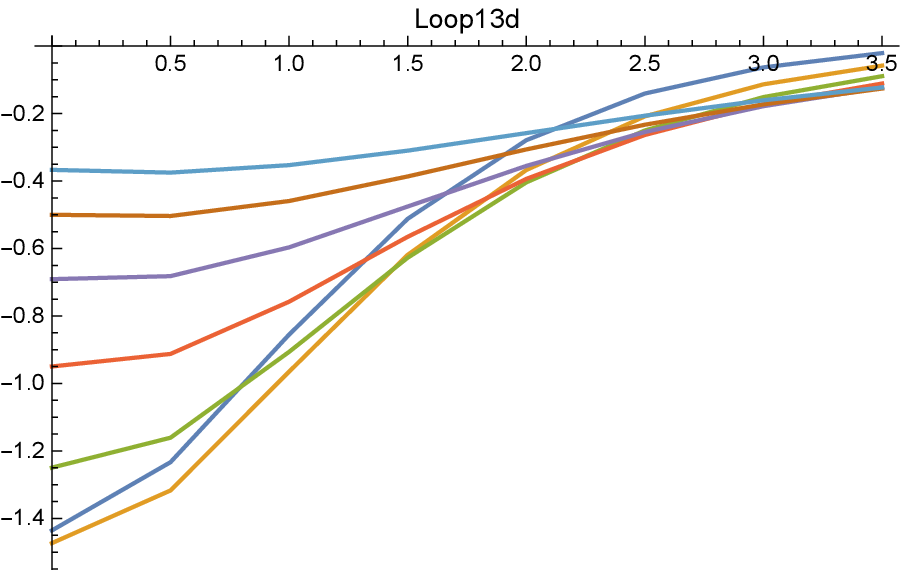}
\end{center}
\end{minipage}
\caption{Trace of $d{\bf S}(L13[u_1,u_2])$ for $\Delta{u_i}=1 $(left) and $\Delta{u_i}=\frac{1}{2} $ (right). ($i=1,2)$ }
\label{L13tr}
\end{figure*}

The traces of the matrix in the case of $Loop13$ are shown in Fig.\ref{L13tr}.

\begin{figure*}[htb]
\begin{minipage}[b]{0.47\linewidth}
\begin{center}
\includegraphics[width=6cm,angle=0,clip]{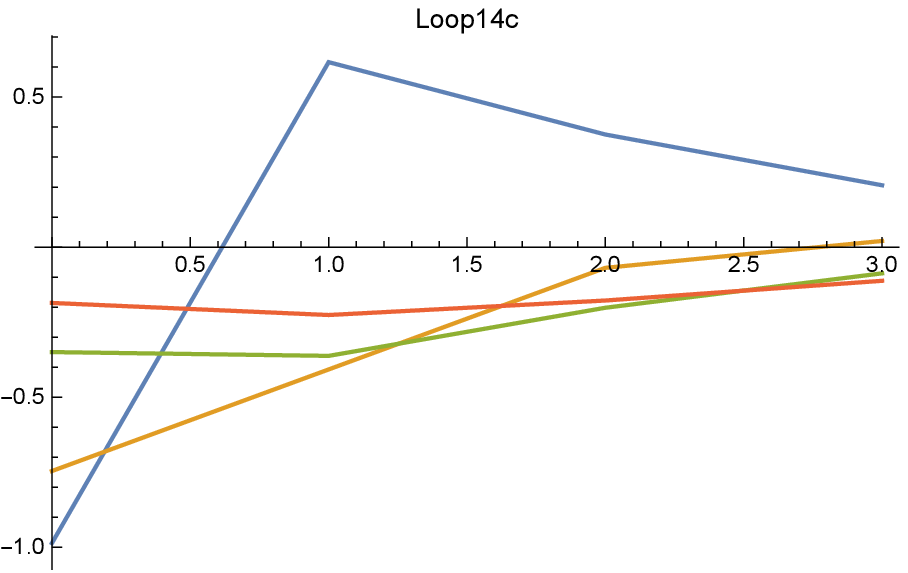}
\end{center}
\end{minipage}
\hfill
\begin{minipage}[b]{0.47\linewidth}
\begin{center}
\includegraphics[width=6cm,angle=0,clip]{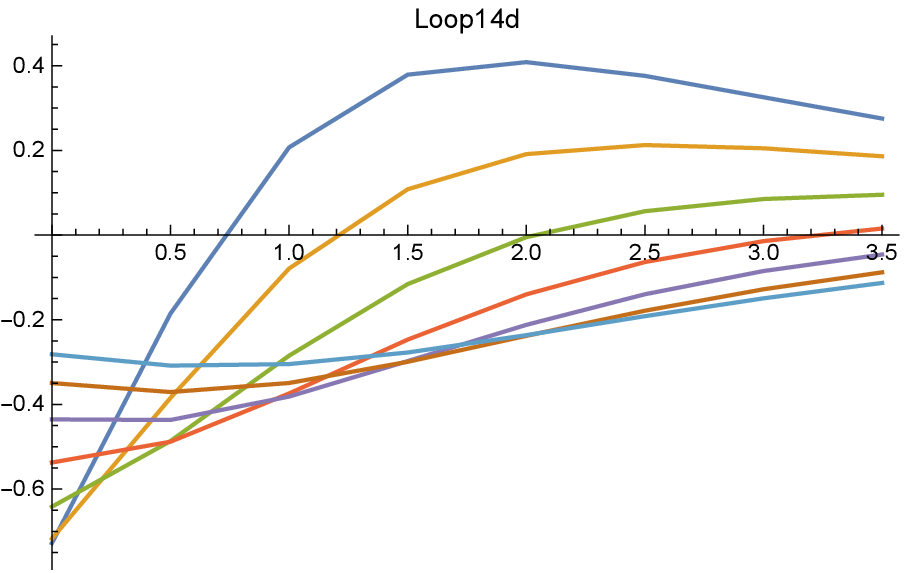}
\end{center}
\end{minipage}
\caption{Trace of $d{\bf S}(L14[u_1,u_2])$ for $\Delta{u_i}=1 $(left) and $\Delta{u_i}=\frac{1}{2} $ (right). ($i=1,2)$ }
\label{L14tr}
\end{figure*}

The traces of the matrix in the case of $Loop14$ are shown in Fig.\ref{L14tr}.

\begin{figure*}[htb]
\begin{minipage}[b]{0.47\linewidth}
\begin{center}
\includegraphics[width=6cm,angle=0,clip]{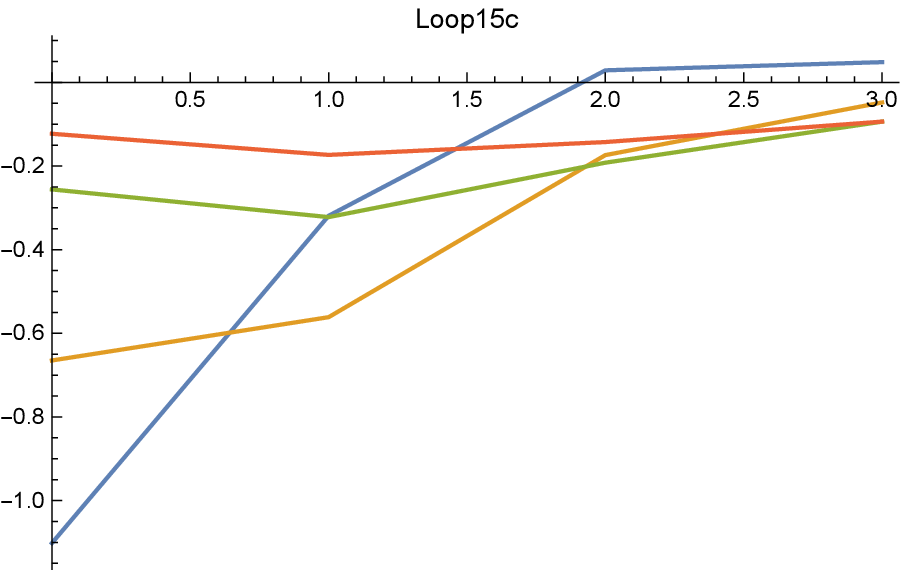}
\end{center}
\end{minipage}
\hfill
\begin{minipage}[b]{0.47\linewidth}
\begin{center}
\includegraphics[width=6cm,angle=0,clip]{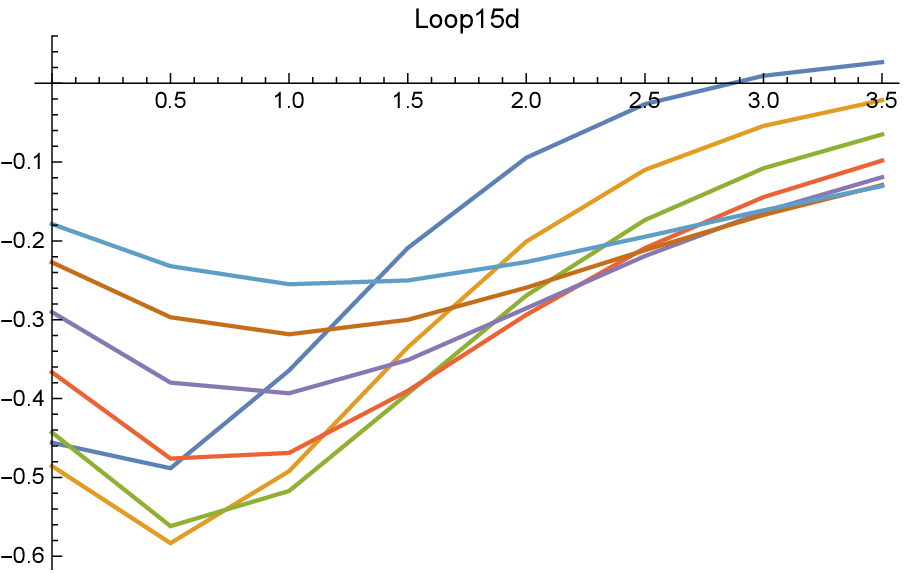}
\end{center}
\end{minipage}
\caption{Trace of $d{\bf S}(L15[u_1,u_2])$ for $\Delta{u_i}=1$(left) and $\Delta{u_i}=\frac{1}{2} $ (right). ($i=1,2)$ }
\label{L15tr}
\end{figure*}

\newpage
The traces of the matrix in the case of $Loop15$ are shown in Fig.\ref{L15tr}.

\begin{figure*}[htb]
\begin{minipage}[b]{0.47\linewidth}
\begin{center}
\includegraphics[width=6cm,angle=0,clip]{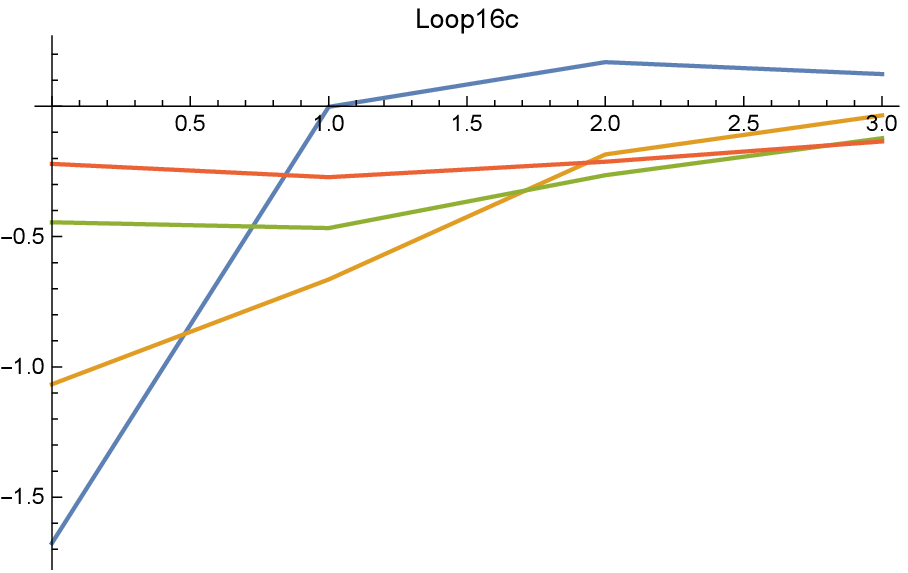}
\end{center}
\end{minipage}
\hfill
\begin{minipage}[b]{0.47\linewidth}
\begin{center}
\includegraphics[width=6cm,angle=0,clip]{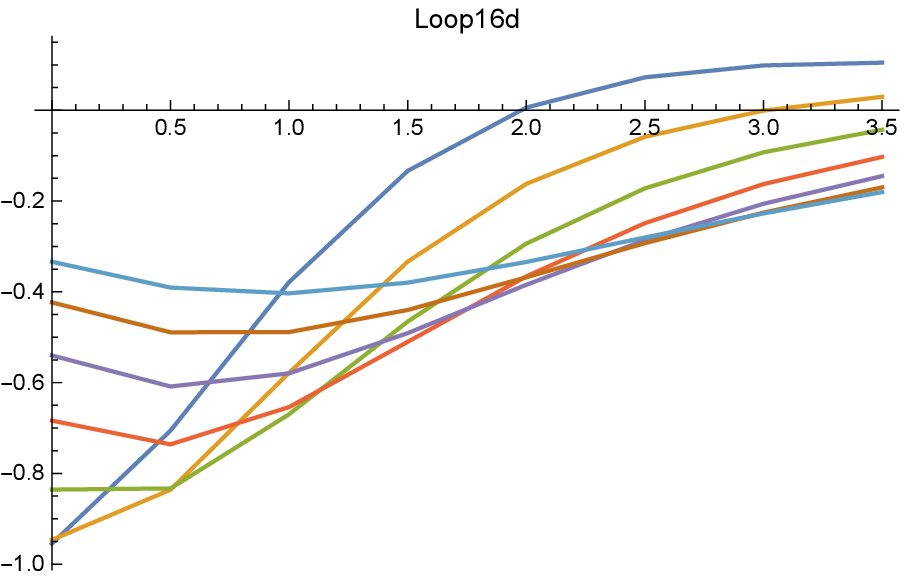}
\end{center}
\end{minipage}
\caption{Trace of $d{\bf S}(L16[u_1,u_2])$ for $\Delta{u_i}=1 $(left) and $\Delta{u_i}=\frac{1}{2} $ (right). ($i=1,2)$ }
\label{L16tr}
\end{figure*}

The traces of the matrix in the case of $Loop16$ are shown in Fig.\ref{L16tr}.

\begin{figure*}[htb]
\begin{minipage}[b]{0.47\linewidth}
\begin{center}
\includegraphics[width=6cm,angle=0,clip]{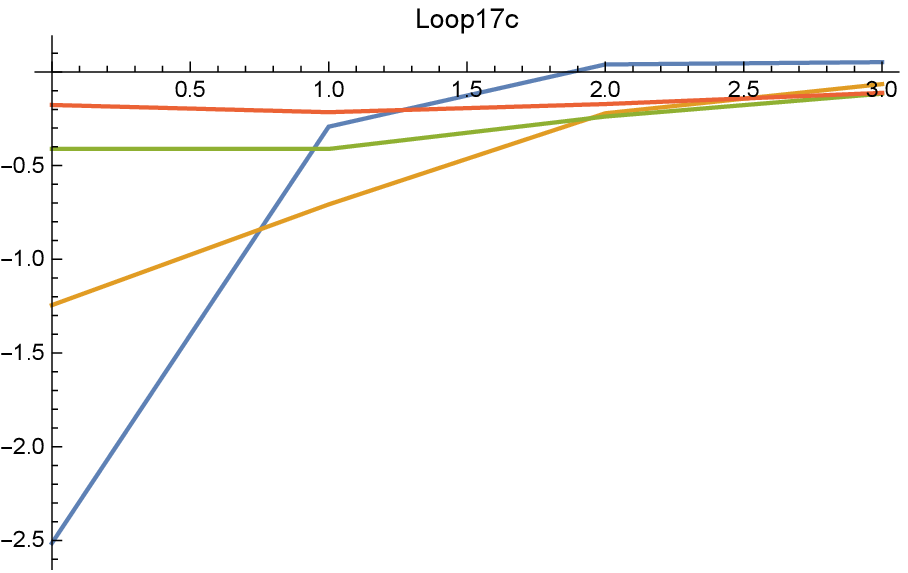}
\end{center}
\end{minipage}
\hfill
\begin{minipage}[b]{0.47\linewidth}
\begin{center}
\includegraphics[width=6cm,angle=0,clip]{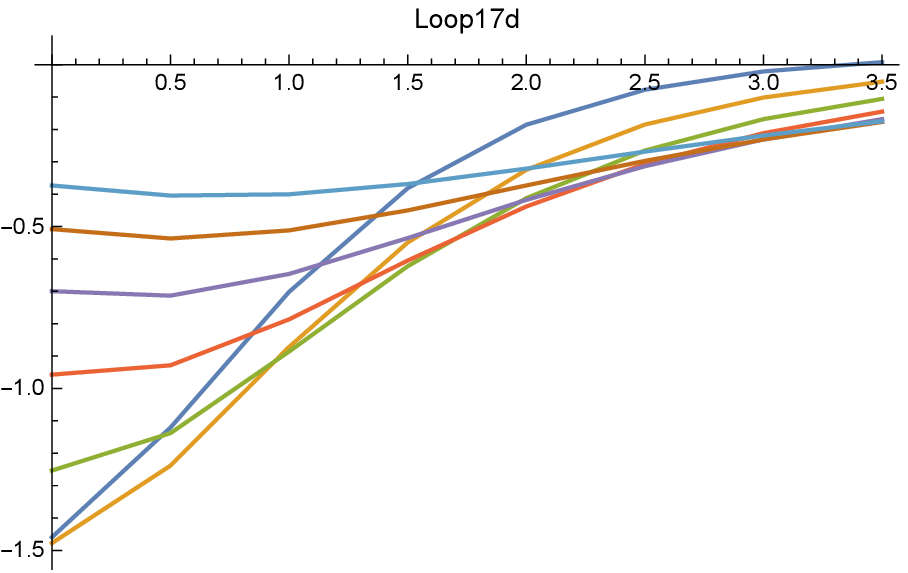}
\end{center}
\end{minipage}
\caption{Trace of $d{\bf S}(L17[u_1,u_2])$ for $\Delta{u_i}=1 $(left) and $\Delta{u_i}=\frac{1}{2} $(right). ($i=1,2)$ }
\label{L17tr}
\end{figure*}

The traces of the matrix in the case of $Loop17$ are shown in Fig.\ref{L17tr}.

\begin{figure*}[htb]
\begin{minipage}[b]{0.47\linewidth}
\begin{center}
\includegraphics[width=6cm,angle=0,clip]{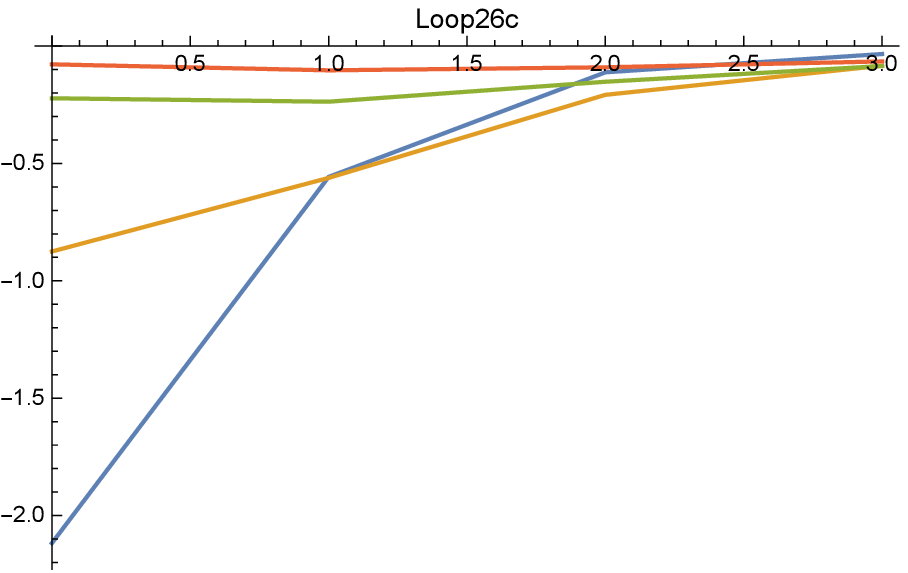}
\end{center}
\end{minipage}
\hfill
\begin{minipage}[b]{0.47\linewidth}
\begin{center}
\includegraphics[width=6cm,angle=0,clip]{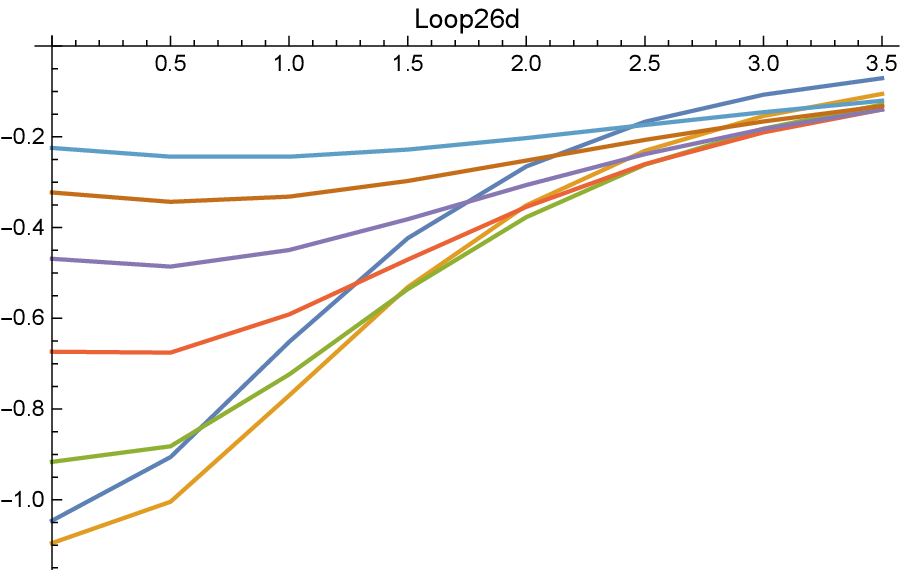}
\end{center}
\end{minipage}
\caption{Trace of $d{\bf S}(L26[u_1,u_2])$ for $\Delta{u_i}=1 $(left) and $\Delta{u_i}=\frac{1}{2} $(right). ($i=1,2)$ }
\label{L26tr}
\end{figure*}

\newpage
The traces of the matrix in the case of $Loop26$ are shown in Fig.\ref{L26tr}.

\begin{figure*}[htb]
\begin{minipage}[b]{0.47\linewidth}
\begin{center}
\includegraphics[width=6cm,angle=0,clip]{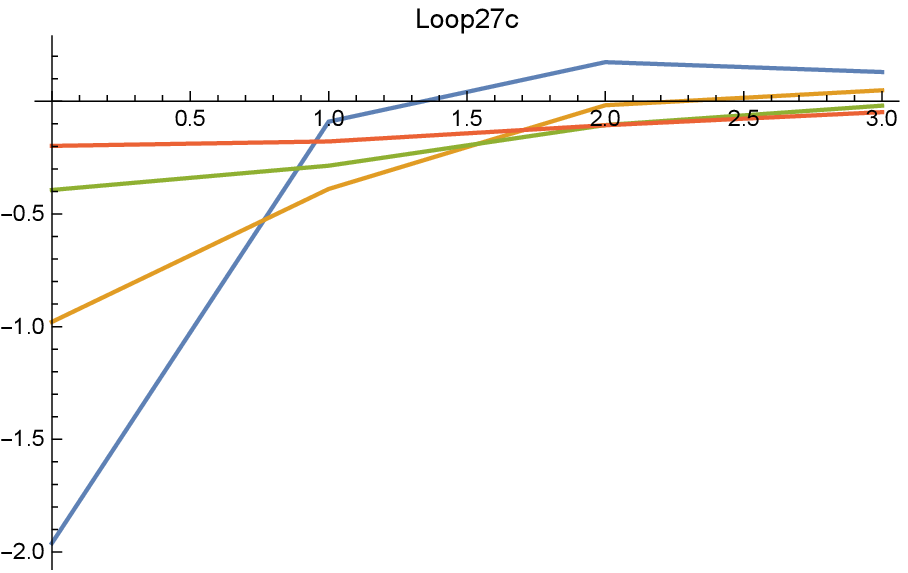}
\end{center}
\end{minipage}
\hfill
\begin{minipage}[b]{0.47\linewidth}
\begin{center}
\includegraphics[width=6cm,angle=0,clip]{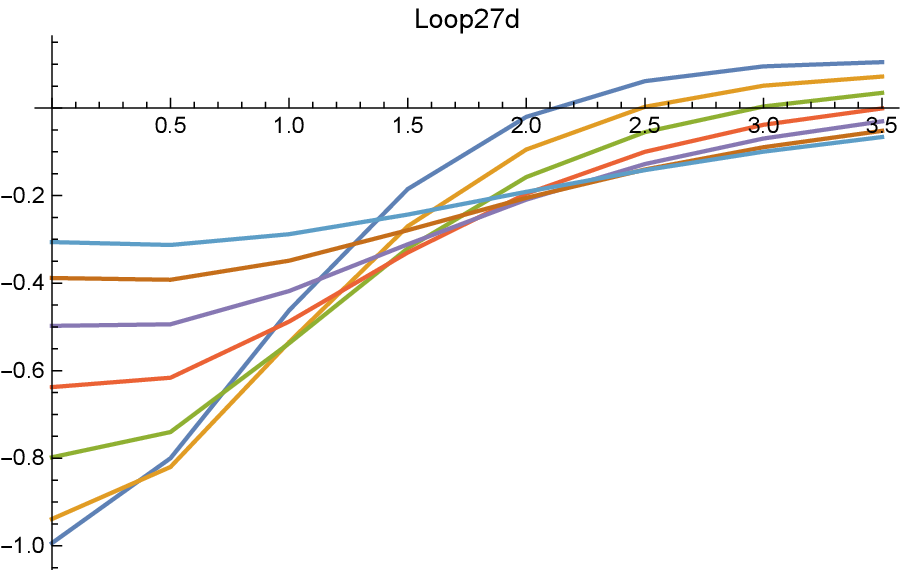}
\end{center}
\end{minipage}
\caption{Trace of $d{\bf S}(L27[u_1,u_2])$ for $\Delta{u_i}=1 $(left) and $\Delta{u_i}=\frac{1}{2} $(right). ($i=1,2)$ }
\label{L27tr}
\end{figure*}

The traces of the matrix in the case of $Loop27$ are shown in Fig.\ref{L27tr}.

 \section{Discussion and perspective}
We expected that correlation of phonons and its time-reversed phonons propagating on a $2D$ plane can be simulated by a model of Bosonic
quasiparticle propagating in the Fermionic sea of Weyl spinors.  

In an exploratory analysis using Clifford algebra, we observed that as the lattice spacing $a$ is halved, eigenvalues of the FP Wilson action and the trace of the matrix representing links of Loops surrounding each FP Wilson action are
reduced.  Monte-Carlo simulations using a small lattice constant $a$ will allow fixing the optimal Wilson action or the Polyakov action as a linear combination of the FP actions with correction terms derived by the renormalization group. 

There are works on the study of Quantum spin systems associated with time translations and space translations of lattice spaces\cite{Robinson68,LR72}. 
Solitons are integrable systems induced by nonlinear interactions. Detailed comparison between experiments and simulation will clarify the Kubo-Martin-Schwinger boundary condition. 

We have a plan of reducing $a$ and perform the renormalization group analysis using supercomputers which allow parallel computations.  
 Whether one can connects the chiral anomaly and grvitation alanomaly by simulating the ultrasonic waves remanis as a future study.

\begin{acknowledgements}
I thank Dr. Serge DosSantos at INSA for valuable discussion and Prof. M. Arai for supports. Thanks are also due to the CMC of Osaka University for allowing use of super computers there, which were developed by CMC and Tokyo Institute of Technology.
\end{acknowledgements}

\end{document}